\documentclass[12pt]{article}
\newlength{\abstractwidth}
\usepackage{a4}
\usepackage{amsmath}
\usepackage{amssymb}
\usepackage{ulem}

\usepackage[dvips]{graphicx}
\usepackage{color}
\def\beq{\begin{equation}}
\def\eeq{\end{equation}}
\def\beqa{\begin{eqnarray}}
\def\eeqa{\end{eqnarray}}
\def\n{\nonumber \\}

\def\rc{\ensuremath{{\cal R}}}
\def\tc{\ensuremath{{\cal T}}}
\def\sech{{\rm sech}}

\newcommand {\be}{\begin{eqnarray}}
\newcommand{\ee}{\end{eqnarray}}
\catcode`\@=11
\@addtoreset{equation}{section}

\catcode`@=12
\relax

\begin{document}

\begin{flushright}
{SAGA-HE-289}\\
{KEK-TH-2000}
\end{flushright}
\vskip 0.5 truecm

\begin{center}
{\Large{\bf
Vacuum fluctuations in an ancestor vacuum:\\
A possible dark energy candidate
}}\\
\vskip 1cm

Hajime Aoki$^{a}$\footnote{haoki@cc.saga-u.ac.jp},
 Satoshi Iso$^{b}$\footnote{iso@post.kek.jp}, Da-Shin Lee$^{c}\footnote{dslee@gms.ndhu.edu.tw}$,\\
 Yasuhiro Sekino$^{d}$\footnote{ysekino@la.takushoku-u.ac.jp}  and Chen-Pin Yeh$^{c}$\footnote{chenpinyeh@gmail.com} \\

\vskip 0.5cm

$^a${\it Department of Physics, Saga University, Saga 840-8502,
Japan}\\
$^b${\it Theory Center,
 High Energy Accelerator Research Organization (KEK),\\
 and Graduate University for Advanced Studies (SOKENDAI),\\
Ibaraki 305-0801, Japan}\\
 $^c${\it Department of Physics, National Dong-Hwa University,\\
 Hualien 97401, Taiwan, R.O.C.}\\
 $^e$
{\it  Department of Liberal Arts and Sciences,
Faculty of Engineering,\\
Takushoku University, Tokyo 193-0985, Japan}
\end{center}

\vskip 1cm

\begin{abstract}
 We consider an open universe created by bubble nucleation, and study
 possible effects of our ``ancestor vacuum,'' a de Sitter space
 in which bubble nucleation occurred, on the present universe.
 We compute vacuum expectation values of the energy-momentum tensor
 for a minimally coupled scalar field, carefully taking into account the
 effect of the ancestor vacuum by the Euclidean prescription.
 We pay particular attention to the so-called supercurvature mode,
a non-normalizable mode on a spatial slice of the open universe,
which has been known to exist for sufficiently light fields.
 This mode decays in time most slowly, and may leave
 residual effects of the ancestor vacuum, potentially observable
 in the present universe.
 We point out that the vacuum energy of the quantum field can be
 regarded as dark energy if mass of the field is of order the
 present Hubble parameter or smaller. We obtain preliminary
 results for the dark energy equation of state $w(z)$ as a
 function of the redshift.

\end{abstract}

\baselineskip=17.63pt

\newpage
\section{Introduction}
\label{sec:introduction}
\setcounter{footnote}{0}
\setcounter{equation}{0}

In a theory with multiple vacua, nucleation of bubbles
of a true vacuum can occur due to quantum tunneling.
If such a theory is coupled to gravity, bubble nucleation
provides a mechanism for the creation of an
FLRW(Fiedmann-Lema\^{\i}tre-Robertson-Walker)
universe. Superstring theory
is expected to have
a large number of metastable vacua~\cite{BoussoPolchinski,
Douglas, SusskindLandscape} including positive energy de Sitter
vacua~\cite{KKLT, BBCQ},
according to the proposal of the ``string landscape''
(see e.g., \cite{landscape2} for a review).
Although the existence of such vacua has not been
proven yet, we consider bubble nucleation to be a viable
mechanism to determine initial conditions for
our universe.

Understanding of the characteristic features of
the universe created by bubble nucleation is
of great importance. One of such features is
negative spatial curvature~\cite{KlebanSchillo}.
The Coleman-De Luccia instanton~\cite{CDL},
which is a semi-classical description of the
creation and evolution of a bubble, shows that
the universe inside the bubble should have
negative curvature.
Even though our universe is known to be flat within the
margin of error, it is logically possible that there is
a finite radius $R_c$ of negative curvature. The bound is
roughly $R_c\gtrsim 10 H_0^{-1}$ where
$H_0$ is the current Hubble parameter~\cite{PlanckCosm}.

Another feature would be the possible signatures in
the cosmic
microwave background radiation (CMB).
The spectrum of the CMB temperature fluctuations is
consistent with the nearly scale-invariant spectrum of
primordial fluctuations predicted by inflation~\cite{PlanckInfl}.
However, if the number of e-folds of inflation is finite,
we might be able to see deviations
from scale invariance due to the evolution of the universe
before inflation. Such effects are expected to
affect the low-$\ell$ (angular momentum) modes of
the power spectrum. The CMB spectrum in the universe
created by bubble nucleation has been understood in the
1990's\footnote{Although it was often assumed at that time
that the energy density from the spatial curvature is
as large as what we now know to be dark energy
(since the observational evidence for the cosmic acceleration
has not been established yet),
essential features of the spectrum has been
understood.}~\cite{SasakiTanakaYamamoto1, SasakiTanakaYamamoto2,
SasakiTanakaYamamoto3, SasakiTanakaGraviton,
GarrigaMontes1, GarrigaMontes2, LindeSasakiTanaka}.
The studies on the CMB spectrum after the advent of
string landscape include \cite{FKMS1, FKMS2,
Yamauchi,  BoussoHarlow1, BoussoHarlow2}.
Although no signature of bubble nucleation has been found
in the CMB yet, theoretical study for seeking such a signature
is undoubtedly important.

 In this paper, we will focus on a third possible feature,
 related but different from the above two.
 We will consider how the
  quantum fluctuations generated before bubble nucleation
  affect the vacuum energy in the present universe.
The universe created by bubble nucleation is
surrounded by a parent de Sitter space,
which we call the ancestor vacuum (See Fig.~\ref{Fig-Penrose}).
The Hubble parameter $H_A$ for the ancestor vacuum is
typically larger than the value $H_I$ for inflation
after tunneling. On dimensional grounds, large fluctuations
will be generated in the ancestor vacuum.
In this paper, we will compute the vacuum expectation value of
the energy-momentum tensor of a free scalar field,
carefully taking into account the
effect of the ancestor vacuum.

To find the vacuum expectation value of the energy-momentum
tensor,
which is quadratic in the quantum field, we compute the
two-point functions of the field and take the
coincident-point limit. The two-point
functions at early
time (in our universe) are obtained by the Euclidean prescription,
and their subsequent time-evolution can be studied using the equation of
motion for the scalar field.

We will be particularly interested in the contribution from
a special mode of fluctuations,  the so-called
``supercurvature mode\footnote{It is difficult to find a
situation in which the supercurvature mode affects the CMB.
Fluctuations
of the tunneling field should have large mass in order
for the Coleman-De Luccia instanton to exist~\cite{CDL},
thus they will not have a supercurvature mode.
Multi-field models (the tunneling and inflaton fields
being different) have a potential problem as
pointed out in \cite{SasakiTanakaTwofield},
but there is a recent attempt~\cite{LiddleCortes, Kanno}
at explaining a ``dipolar anomaly'' in the CMB using
a supercurvature mode in curvaton models.
Gravitons (tensor modes) are massless, but their supercurvature modes
are pure gauge~\cite{SasakiTanakaGraviton}.},'' originally found
by M.~Sasaki, T.~Tanaka and K.~Yamamoto~\cite{SasakiTanakaYamamoto1, SasakiTanakaYamamoto2,
SasakiTanakaYamamoto3}~\footnote{The supercurvature mode has played an
important role in an attempt to construct holographic dual for
universe created by bubble nucleation~\cite{FSSY, SekinoSusskind}.
In these papers, the supercurvature
mode was called the non-normalizable mode.}.
This mode decays more slowly than $e^{-R}$ at large $R$
where $R$ is the geodesic distance, and is not normalizable
on the spatial slice $H^{3}$
in the open universe. The supercurvature mode appears if the mass of
the scalar field in the ancestor vacuum is small enough.
For example, for exactly massless fields, two-point functions
remain finite even when two points are infinitely separated
on $H^3$. Heuristically, such fluctuations can be considered as
the super-horizon fluctuations in the ancestor (de Sitter)
vacuum, seen from the inside of the bubble~\cite{SekinoSusskind}.
Mathematical reason that a non-normalizable mode can exist
in the open universe is that the spatial slice
$H^3$ is not a global Cauchy surface
(see Fig.~\ref{Fig-Penrose}). To quantize the fluctuations,
one needs to take a complete set of normalizable modes
on a Cauchy surface, such as the surface on the
horizonal brown line in Fig.~\ref{Fig-Penrose}.
The supercurvature mode appears as a result of analytic continuation
of the correlator to the open universe, as we will review
in Section~3.

Supercurvature modes decay more slowly than normalizable modes,
not only in space but also in time. Thus, it may have
a chance to affect our current universe. In previous
papers by some of the present
authors~\cite{AIS, AI}\footnote{See also
\cite{Ringeval,Glavan:2013mra,Glavan:2014uga,Glavan:2015cut,Glavan:2017jye}
for related work.},
the  evolution of vacuum fluctuations generated
during and before inflation has been studied, without taking
bubble nucleation into account. The emergence of the supercurvature
mode is an interesting phenomenon, special to the universe created by
bubble nucleation.

As long as the mass of the scalar field is
smaller than the Hubble parameter in our universe, the field value
for the supercurvature mode remains nearly constant. This is the
well-known freezing of the super-horizon fluctuations;
supercurvature modes can always be considered to be outside the
horizon. The energy-momentum tensor for this mode
behaves similarly to that for cosmological constant. After the Hubble
parameter decreases below the mass, the field
begins to oscillate, and its energy
decays\footnote{The energy density of a homogeneous field oscillating
in time (averaged over the period)
decays at the same rate as matter energy density
as universe expands, $\rho\sim a^{-3}$.}.
If there
is a field with mass of order the present Hubble parameter
$H_0$ or smaller,
the
supercurvature mode of this field will be essentially frozen until
today. This gives us an interesting possibility for the
realization of dark energy.

The primary purpose of this paper is to explain how to
calculate the vacuum expectation value of
the energy-momentum tensor in the background with
bubble nucleation.
To the best of our knowledge, this calculation has not been
done before. Another purpose is to show that
dark energy can be obtained as a contribution
from the supercurvature mode to the vacuum energy.

We consider the fluctuations of a minimally coupled
scalar field $\phi$,
which is different from the tunneling field $\Phi$.
The field $\phi$ does not have the vacuum expectation value,
and is treated as a free field in the curved spacetime.
We assume the mass $m_A$ of $\phi$ in the ancestor vacuum
to be sufficiently small relative to the Hubble parameter,
$m_A\ll H_A$, so that a supercurvature mode exists.
We allow a possibility that mass $m_0$ of $\phi$
in the true vacuum is different from $m_A$. (This can be
realized e.g.,
if there is a coupling of the form $\Phi^2\phi^2$, and
the  expectation  value of $\Phi$ is large in the false vacuum and
small in the true vacuum.) It would be
natural to assume $m_0\ll m_A$, since the energy scale
in the true vacuum will be typically lower than in the
false vacuum.

The order of magnitude of the vacuum energy (derived
rigorously below) can be estimated as follows.
The expectation value of the field squared
$\langle \phi^2 \rangle$ in the ancestor vacuum
goes as $\langle \phi^2 \rangle\sim H_A^4/m_A^2$.
This is essentially the same as the well-known expression
in pure de Sitter space,
which becomes infinitely large in the massless
and infinite e-folds limit
(see e.g., \cite{BirrellDavies, Linde})\footnote{In the case of
an inflation with a {\it finite} e-folds $N$,
we instead have $\langle \phi^2 \rangle\sim N H^2$,
where $H$ is the Hubble parameter for inflation.
See, e.g.,\cite{AIS,AI} and references therein.}.
If we assume $m_0\lesssim H_0$, the field
value $\langle \phi^2 \rangle$ is nearly frozen
until now, apart from  its weak time dependence due to
the non-zero $m_A$ that may play an important role
in observationally distinguishing our mechanism from others.
The dominant part of the energy-momentum
tensor for the field $\phi$ is the mass term,
$\rho \sim m_0^2\langle \phi^2\rangle \sim (m_0/m_A)^2H_A^4$.
Then, it is possible to make this of the same order as dark energy,
$\rho\sim H_0^2 M_P^2$ where $M_P$ is the (reduced) Planck mass.
For instance, if there is a field with $m_0\sim H_0$, we need
$M_P/H_A\sim H_A/m_A$ (i.e., $H_A$ being the geometric mean
 of $M_P$ and $m_A$), which does not seem
particularly difficult to satisfy.

In this scenario we need an ultra-light field with
mass $m_0\sim H_0\sim 10^{-33}{\rm eV}$. Thus, this may not
be regarded as a ``natural'' solution to the cosmological constant
problem~\cite{WeinbergCC}. Nevertheless, we believe the detailed study
is worthwhile,
especially in view of the proposal of the ``string axiverse,''
which states that
there is a large number of axion-like
particles\footnote{The statement of the string axiverse is that
if the QCD axion, responsible for the solution of the strong CP
problem, exists in string theory, one should also expect many
other axion-like light particles.}
with mass ranging down to $m_0\sim H_0$~\cite{Axiverse}.

The idea of realizing dark energy as the vacuum energy of
an ultra-light field is not essentially new. See e.g.,
\cite{Axiverse, Hill, Marsh, Ringeval, AI, Glavan:2015cut,
DSLee} for previous works. Summaries of this topic can be
found in review articles~\cite{Copeland, MarshReview}.
The effect of bubble nucleation has not been
considered previously. At the end of this paper,
we will comment on the possible observable effects, which
could be regarded as a signature of our mechanism.

This paper is organized as follows. In Section 2, we
review the Coleman-De Luccia instanton, which
describes the nucleation and the evolution of a bubble
in de Sitter space (the ancestor vacuum). In Section 3,
we review the calculation of correlation functions
of a scalar field based
on analytic continuation from the Euclidean space. Using this
prescription, we obtain the correlators in the early-time
limit in the open universe.
In Section 4, we compute the expectation value of the
energy-momentum tensor by taking the coincident-point
limit of the two-point function obtained in the previous
section. In particular, we study the mass term
in the energy-momentum tensor in the limit of small mass.
In Section 5, we consider the evolution of the energy density
in the open universe.
We first obtain the scale factor for the open FLRW universe
in the eras of curvature domination, inflation, radiation and
matter domination.
We then solve the equation of motion for the scalar field
in each era, and
find the wave function by smoothly connecting the
solution to the wave function in the early-time limit.
Using this wave function, we obtain the expectation
value of the energy-momentum tensor at late times.
In Section~6, we will summarize our results and
comment on the directions for future work.
In Appendix~\ref{sec-modes}, we describe the harmonics
on $H^3.$
In Appendix~\ref{sec-puredS}, we calculate
$\langle \phi^2\rangle$ in pure de Sitter space using
our method, and show that it agrees with the
known value obtained by standard techniques.
In Appendix~\ref{sec-radmatt}, we find
the scale factor for an open universe which
contains both matter and radiation.
In Appendix~\ref{sec-matwf}, we present the
details on the matching of the wave functions of
the scalar field across different eras.

\section{The CDL geometry}
\label{sec-background}

As a model of bubble nucleation, we consider a scalar field $\Phi$
with a potential which has two minima, like the one depicted in
Fig.~\ref{Fig-Penrose}. This field $\Phi$ goes through tunneling
from the false vacuum to the true vacuum ($\Phi$ should not be
confused with the field $\phi$, which will be introduced later and
has the zero expectation value). For phenomenological reasons, we need
inflation after tunneling. Thus, we assume the potential has a
plateau region on the true vacuum side, on which slow-roll
inflation occurs.
We assume the potential at the true vacuum is zero, 
since we expect the present cosmological constant
(dark energy) to be solely due to 
the vacuum energy of a quantum
field\footnote{The study of the back reaction to the geometry
from the quantum vacuum energy thus obtained
is left for future work.}.

In the theory with gravity, the geometry for the false vacuum with
a positive vacuum energy is de Sitter space.
We assume that there is a global surface without bubbles
at some time in de Sitter space\footnote{Without this assumption,
the whole de Sitter space would be swallowed by bubbles
nucleated in the past.}.
Then, bubbles of true vacuum will form inside
the ``ancestor vacuum'' (parent de Sitter space).
This process is described
by the Coleman-De Luccia (CDL) instanton~\cite{CDL}.

Even though our goal is to explore the evolution of the
universe that results from
the scalar potential in Fig.~\ref{Fig-Penrose}, we first study
the thin-wall limit, in which the transition
occurs sharply from a false vacuum with positive vacuum energy
to a true vacuum with {\it zero} vacuum energy.
The purpose of this analysis is to understand the
{\it early}-time behavior in the FLRW universe inside the bubble.
The universe at early times is dominated by the spatial curvature,
thus the constant vacuum energy is unimportant and can be
set to zero.
In principle, one should be able to
obtain the entire evolution of our universe
by analytic continuation
from Euclidean. However, for that purpose one needs to
know the corresponding Euclidean metric
to an infinite precision, which is clearly
an intractable task. Thus we will use analytic continuation
to obtain the early-time behavior only, and solve
the equation of motion for the scalar field
to obtain the subsequent evolution in the
FLRW universe.
The latter analysis will be done in Section 5.

\subsection{Causal structure}

\begin{figure}[!htb]
\center
\includegraphics [scale=.3]
 {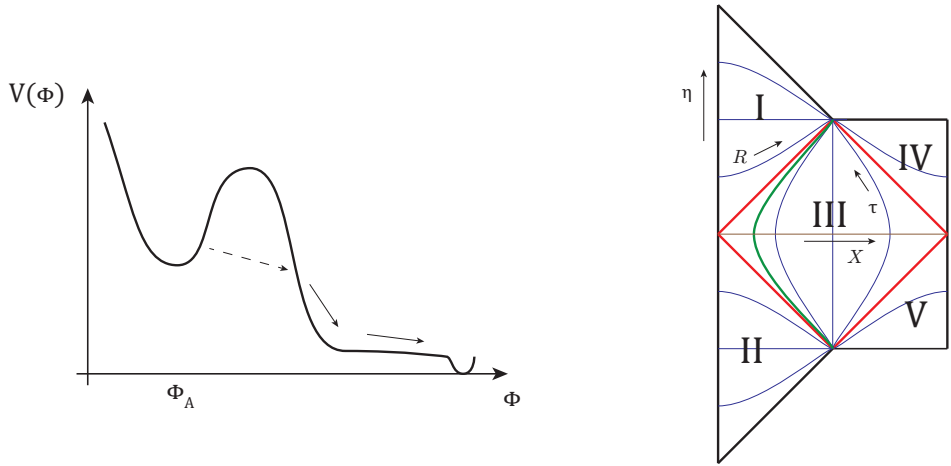} \caption{Left panel: The potential for the field $\Phi$.
 The local minimum at $\Phi=\Phi_A$ is the false vacuum, whose
 geometry is de Sitter space
 (the ancestor vacuum). On the true vacuum side, it is assumed
 that a plateau region exists, on which slow-roll inflation occurs.
 The true minimum of the potential is taken to be zero, since we
 expect the present cosmological constant to be realized as
 the vacuum energy of a quantum field on this background.
 Right panel: Penrose diagram for the Coleman-De Luccia geometry.
 The spacetime is divided into
 five regions by the red lines. Region~I is an open FLRW universe.
 The green line represents the bubble wall. On its left (right) is
 the true (ancestor) vacuum.
 The blue curves indicate orbits of the $SO(3,1)$ symmetry.
 The directions of the coordinates ($\eta$, $R$ for Region I;
 $\tau$, $X$ for Region III) are indicated by arrows. Region~I
 is drawn with the future null infinity, because the present
 cosmological constant will relax to zero in the future
if it is due to the vacuum energy of a quantum field.}
\label{Fig-Penrose}
\end{figure}

The Penrose diagram for the spacetime containing one bubble is
depicted in Fig.~\ref{Fig-Penrose}. Our Universe is an open FLRW
universe inside the bubble (Region I). The beginning of the FLRW
time is the 45-degree line shown in red.
Even though the scale factor vanishes there, it is merely a
coordinate singularity.

The constant-time slices in Region I are 3-hyperboloids, $H^3$,
represented by blue lines.
The bubble wall is represented by the green line
in Region III.  On the right of the
domain wall is the ancestor vacuum.
We have depicted the time-reversal symmetric Penrose diagram
to make the symmetry orbits clearer,
but the lower half part of the diagram should be considered
to be unphysical, and replaced by pure de Sitter
space.

The Penrose diagram in Fig.~\ref{Fig-Penrose}, which has
null future infinity in Region I, is for the case of
the zero final cosmological constant (c.c.). We do not know
whether the present c.c.\ (dark energy) persists into the
infinite future. If dark energy is due to the
vacuum energy of a quantum
field as proposed later in this paper, it will relax to zero
in the future, thus we draw the Penrose diagram for this case.
If the final c.c.\ is positive, the 45 degree line for the
null infinity is replaced by a more horizontal curve
which represents spacelike infinity.

The whole geometry and the configuration of the field
$\Phi$ have the $SO(3,1)$ symmetry. The surfaces of constant
$\Phi$ are the slices on which the $SO(3,1)$ symmetry
acts as isometries. In Region III, these surfaces are
timelike, and are (2+1) dimensional de Sitter spaces.
In particular, the world volume of the bubble has this
symmetry. The bubble wall is the ``vacuum domain wall'',
which has no
structure, and is invariant under the Lorentz boost.
On the other hand, in Region I the $SO(3,1)$
symmetric surfaces are spacelike, $H^3$.
In the thin-wall limit with the zero c.c., Region I is
nothing but  part of Minkowski space in the open
slicing (known as the Milne universe).

  \subsection{Euclidean metric and its
  analytic continuation}
The geometry containing one bubble of the true vacuum
is obtained by analytic continuation from the Euclidean
geometry (Coleman-De Luccia [CDL] instanton), which is
believed to contribute dominantly to the path
integral of quantum gravity.

Let us first describe the Euclidean geometry. The Euclidean
version of the 3+1 dimensional
de Sitter space is a sphere $S^4$, while the
CDL instanton is a deformed sphere described by the
metric of the following form,
\begin{equation}
 ds_E^2=a^2(X)\left(dX^2+d\theta^2+\sin^2\theta d\Omega_2^2
\right).
\label{Euclidmetric}
\end{equation}
This geometry is topologically $S^4$, but is deformed in the
$X$ direction. It has an $S^3$ factor parametrized by
$\theta$ and $\Omega_2$, so it preserves the $SO(4)$ subgroup
of $SO(5)$. The scale factor $a(X)$ behaves
as $a(X)=c_{\pm} e^{\pm X}$ with some constant $c_{\pm}$
for $X\to \mp \infty$ due to the fact that the geometry is smooth
(i.e., locally flat) at both ends.
We will mostly consider the case where the true vacuum has
zero c.c., since this is sufficient for the study of the early
time limit in the open FLRW universe (as explained below).
In the thin-wall limit for the true vacuum with a zero c.c.,
the geometry is $S^4$ patched with
a flat disk at the domain wall located at $X=X_0$,
and the scale factor takes the form,
\begin{align}
 a(X)&=H_A^{-1}{e^{X-X_0}\over \cosh X_0}
\quad (X\le X_0), \nonumber\\
&={H_A^{-1}\over \cosh X} \quad (X_0\le X).
\label{Euclida}
\end{align}
The natural length scale associated with this scale factor
is the inverse Hubble parameter $H_A^{-1}$ of the ancestor vacuum.

\begin{figure}[!htb]
\center
\includegraphics [scale=.3]
 {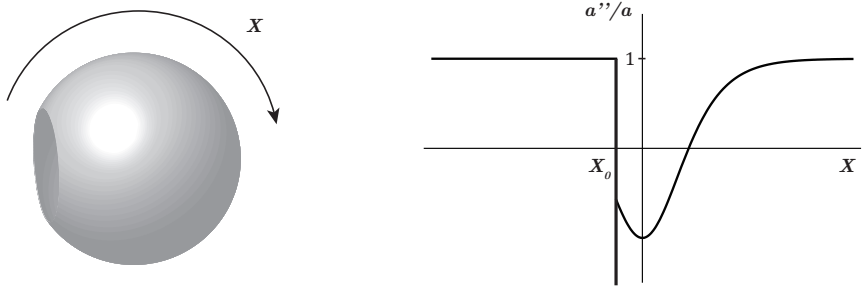}
 \caption{Left panel: The Euclidean geometry for a CDL instanton
 in the thin-wall limit. A flat space for the true
 vacuum (whose vacuum energy is assumed to be zero) is patched
 to a piece of a sphere $S^4$ for the false vacuum
 with a positive vacuum energy. The $S^3$ part is represented
 by $S^1$ in the figure. Right panel: The potential
 $a''/a$  in the massless scalar
 equation of motion \eqref{Schrodinger}
 on a CDL background in the thin-wall limit.
 Asymptotically, $a''/a\to 1$ for $X\to \pm \infty$.
 On the true-vacuum
 side ($X<X_0$), the potential is flat. At the bubble wall
 ($X=X_0$), there is a negative delta function. If mass is non-zero,
 the potential will be lifted by the additional term $+m^2a^2$.}
\label{Fig-DeformedSphere}
\end{figure}

We can analytically continue the geometry
from the reflection symmetric
surface, namely an equator of the $S^3$. With
\begin{equation}
 \theta\to i \tau+{\pi\over 2},
\end{equation}
we obtain the metric covering Region III in
Fig.~\ref{Fig-Penrose},
\begin{equation}
 ds^2=a^2(X)\left( dX^2 -d\tau^2+\cosh^2\tau d\Omega_2^2
\right).
\label{centermetric}
\end{equation}
This has a factor of $(2+1)$-dimensional de Sitter
space, parametrized by $\tau$ and $\Omega_2$. The $SO(4)$
symmetry of the Euclidean space becomes $SO(3,1)$ after
analytic continuation. This symmetry acts as the isometry
group of the (2+1) dimensional de Sitter space.
The coordinate $X$ parametrizes the spatial direction
transverse to these de Sitter spaces.
The $X$-direction for $\tau=0$
is represented by the horizontal line in
Fig.~\ref{Fig-Penrose}.

Region III is not geodesically complete. The spacetime
can be extended past the horizons ($X\to \pm\infty$,
$\tau \to\pm\infty$), represented by the 45 degree red lines
in Fig.~\ref{Fig-Penrose}, which are locally
equivalent to the Rindler horizon. By continuing past
$X\to -\infty$, $\tau\to +\infty$, we obtain an open
FLRW universe (Region I) described by the metric
\begin{equation}
 ds^2=a^2(\eta)\left(-d\eta^2 +dR^2 +\sinh^2 R d\Omega_2^2\right).
\label{FRWmetric}
\end{equation}
The spatial sections are $H^3$, parametrized by $R$ and
$\Omega_2$ with the isometry group $SO(3,1)$.
This metric \eqref{FRWmetric}
is most easily obtained by an analytic continuation,
\begin{equation}
 X\to \eta +{\pi\over 2}i,\quad \theta\to i R,
\label{analyticcont2}
\end{equation}
from the Euclidean metric \eqref{Euclidmetric}.
The scale factor behaves as
\begin{equation}
 a(\eta)\sim  e^\eta
\label{aeta}
\end{equation}
in the early-time limit $\eta\to -\infty$.
In the thin-wall limit with the zero
final c.c., the scale factor is exactly  $a(\eta)=
 const. \times e^\eta$.

One way to understand the relation between the coordinates
in \eqref{centermetric} and \eqref{FRWmetric} is to
note that the geometry near the light cone is locally flat. One
can use the coordinates $\hat{t}$, $\hat{r}$ to cover
the global Minkowski space, $ds_{\rm flat}^2=-d\hat{t}^2+d\hat{r}^2
+\hat{r}^{2}d\Omega_{2}^{2}$. The metric \eqref{centermetric}
with $a(X)=e^{X}$ is obtained by
\begin{equation}
 \hat{t}=e^{X}\sinh\tau, \quad \hat{r}=e^{X}\cosh\tau,
\label{hatt1}
\end{equation}
whereas the metric \eqref{FRWmetric} with $a(\eta)=e^{\eta}$ is obtained by
\begin{equation}
 \hat{t}=e^{\eta}\cosh R, \quad \hat{r}=e^{\eta}\sinh R.
\label{hatt2}
\end{equation}

\subsection{The open FLRW universe}
\label{subsec-FLRW}
The open FLRW universe (Region I) is entirely inside the bubble,
and the CDL instanton sets its initial condition.
The scale factor has to behave as $\eqref{aeta}$ in the early-time
limit.
The evolution afterwards can be found by solving the Friedmann
equation,
\begin{equation}
{(a')^2\over a^4}={1\over 3M_P^2}\rho+{1\over a^2},
\label{Friedmann}
\end{equation}
where the prime denotes the derivative with respect
to the conformal time $\eta$, and $M_P$ is the reduced
Planck scale. The last term in \eqref{Friedmann}
is the contribution from the negative spatial curvature.
The energy density $\rho$ and pressure $p$
satisfy the conservation equation,
\begin{equation}
\rho'+3{a'\over a}(\rho+p)=0.
\label{conservation}
\end{equation}
If the energy density of the universe is dominated by
that of a classical scalar field
$\Phi(\eta)$ which depends only on time, we have
\begin{equation}
\rho={1\over 2a^2}(\Phi')^2+V(\Phi),
\quad p={1\over 2a^2}(\Phi')^2-V(\Phi).
\end{equation}

Just after the beginning of the FLRW time, there
is a curvature dominated era in which the scale factor
is approximately \eqref{aeta}.
(The spacetime curvature is zero in the early-time
limit, and the universe is dominated by the spatial
curvature.)
This era will continue until the vacuum energy
for inflation $\rho=V_I$ starts to dominate
over the contribution from the curvature $V_I\sim M_P^2/a^2$.
The consequences of this curvature dominated era
have been discussed in detail in~\cite{FKMS1}.

Then, slow-roll inflation occurs.
For simplicity, we ignore the gradient of the potential
in the plateau region
in Fig.~\ref{Fig-Penrose}, and assume
that there is a constant energy density $\rho=V_I\equiv 3M_p^2 H_I^2$
from the beginning of the FLRW time until the end of slow-roll inflation.
We will also ignore a possible fast-rolling phase after tunneling
and before slow-roll inflation\footnote{The consequences of
a fast-rolling phase after tunneling
have been discussed e.g., in~\cite{FKMS2,
LindeSasakiTanaka, Yamauchi,
 BoussoHarlow1, BoussoHarlow2}.}.
 We expect these features will give only small corrections to
 our main result.

The solution of \eqref{Friedmann} with $\rho=3M_p^2 H_I^2$ is
\begin{equation}
a(\eta)={H_I^{-1}\over \sinh(-\eta)},
\label{aInf}
\end{equation}
up to a shift of $\eta$.
We could replace $\eta$ in \eqref{aInf} by
$\eta-\tilde{\eta}_1$ where
\begin{equation}
e^{\tilde{\eta}_1}={H_A\over H_I}~(1+e^{2X_0}),
\label{tildeeta}
\end{equation}
so that \eqref{aInf} becomes
$a(\eta)=H_A^{-1}e^{\eta-X_0}/\cosh X_0$
in the early-time limit $\eta\to-\infty$,
to match the normalization of the
Euclidean scale factor \eqref{Euclida},
but we will not do this in the following.
Instead, we will use \eqref{aInf} in the FLRW universe,
and replace $\eta$ by $\eta+\tilde\eta_1$ in the
correlators obtained by analytic continuation from
the Euclidean space.
See eq.~(\ref{FRWcorrelator}) below.

After the slow-roll inflation, reheating occurs, and the
radiation and matter dominated eras
will follow. The evolution through these eras until
the present time will be studied in Section~\ref{sec-evolution}.

\section{Correlation functions}
\label{sec-correlation}

We  now start to calculate the two-point functions in the
early-time limit in the FLRW universe. We will first
find the correlator in the Euclidean space \eqref{Euclidmetric},
and analytically
continue it to the Lorentzian spacetime.
The essential part of the calculation is the
decomposition of the field
into a complete set of states in the $X$ direction (along
the horizontal line in Fig.~\ref{Fig-Penrose}).
For more details of this calculation, see Appendix
of \cite{FSSY}.

\subsection{Calculation in Euclidean space}

We consider the Euclidean CDL geometry, and compute
correlation functions of a minimally
coupled scalar field $\phi$, which is described by
the action
\begin{equation}
 S=\int d^4 x\sqrt{g}~\frac{1}{2}\left(g^{\mu\nu}\partial_\mu\phi
\partial_\nu\phi+m^{2}\phi^{2}\right).
\end{equation}
The field $\phi$ is different from the tunneling field $\Phi$.
We assume $\phi$ to have the zero expectation value,
and can be treated as a free field on the curved
(CDL) background. It is convenient to define a
field $\chi(X, \Omega_3)=a(X) \phi(X, \Omega_3)$,
for which the kinetic term is independent of $a(X)$,
$S=\int dX d\Omega_3 (\partial_{X} \chi)^2+\cdots$.

We would like to obtain the two-point function that satisfies the equation of motion with a
delta-function source,
\begin{equation}
 \left[
-\partial_X^2 +{a''(X)\over a(X)}+m^2 a^2(X)
-\nabla_S^2
\right]\langle \chi(X, \Omega_3)\chi (X', 0)\rangle
=\delta(X-X')\delta^{3}(\Omega_3),
\label{Greenfneq}
\end{equation}
where $\nabla_S^2$ and $\delta^{3}(\Omega_3)$
are the Laplacian and delta function, respectively,
on $S^3$.

We will obtain the correlator as an expansion in
the complete basis in the $X$ direction.
In the Lorentzian geometry, the $X$ direction
lies along a global Cauchy surface (See Fig.~\ref{Fig-Penrose}).
The complete basis is formed
by the eigenfunctions $u^I_k(X)$
of the following differential operator,
\begin{equation}
\left[ -\partial_X^2 +{a''(X)\over a(X)}+m^2a^2(X)\right]u^I_k(X)
=(k^2+1)u^I_k(X),
\label{Schrodinger}
\end{equation}
where $k$ labels the eigenvalue, and $I$ is either
$L$ ($R$) for waves coming from the left (right),
or $B$ for the bound state.
This equation is of the form of the time-independent
Schr\"{o}dinger equation in one dimension.
The potential $a''/a+m^2a^2$ approaches
$+1$ for $X\to \pm \infty$ (see Fig~\ref{Fig-DeformedSphere}).
Thus, the modes for real $k(>0)$
with the eigen-energy
larger than $+1$ are the waves oscillating in $X$.
The orthogonal set consists of the waves coming from
the left $u_k^{L}(X)$,
and those from the right $u_k^{R}(X)$, which satisfy
\begin{align}
u^L_k(X)
 &\to e^{ikX}+\rc (k)e^{-ikX}\quad (X\to -\infty),
\nonumber\\
&\to \tc (k )e^{ikX}\quad (X\to \infty),
\label{uL}
\end{align}
and
\begin{align}
u^R_k(X)
 &\to \tc_R (k)e^{-ikX}\quad (X\to -\infty),\nonumber\\
&\to e^{-ikX}+\rc_R (k)e^{ikX}\quad (X\to \infty).
\label{uR}
\end{align}
$\rc (k)$ and $\tc (k )$ are the reflection and transmission
coefficients for the scattering from the left, and
$\rc_R(k)$ and $\tc_R(k)$ are those for the
scattering from the right. They are related
by $\tc (k )=\tc_R(k)$, $\rc (k)/\rc_R(k)^*=-\tc (k)/\tc (k)^*$.
For negative real $k$, we define
$\rc (-k)=\rc ^*(k)$, $\tc (-k)=\tc ^*(k)$.
From the conservation of the probability current, we have
$\rc (k)\rc ^*(k)+\tc (k )\tc^*(k)=1$.
(See e.g., \cite{barton} for basic facts about
the scattering problem in one dimension.)
These modes satisfy the orthogonality property
\begin{equation}
 \int_{-\infty}^{\infty}dX u^I_k(X)u^{*J}_{k'}(X)
  =2\pi\delta^{IJ}\delta(k-k'),
\end{equation}
where $I, J$ are either $L$ or $R$.

If there are bound states in \eqref{Schrodinger},
one has to include them in the complete set.
Bound states occur at discrete imaginary values
of the momentum $k=k_B$ with ${\rm Im}(k_B)>0$. At
$k=k_B$, both $\rc(k_B)$ and $\tc(k_B)$ have a pole.
This means that $e^{\pm ik_{B}X}$ in $u^{L/R}_k(X)$ is
negligible compared to
$\rc(k_B) e^{\mp i k_{B}X}$ and $\tc(k_B) e^{\pm i k_{B}X}$,
thus
the wave function \eqref{uL} with $k=k_B$ decays at both ends
$X\to \pm \infty$, and is normalizable.

Let us first consider the massless case.  In this case,
one can easily see that
\begin{equation}
 u^B_{k=i}(X)={\cal N} H_A a(X)
\label{uB}
\end{equation}
satisfies \eqref{Schrodinger} with $k=i$ where ${\cal N}$ is a dimensionless constant. An example of the
potential ${a''/a}$ for the Schr\"{o}dinger equation~\eqref{Schrodinger}
is depicted in Fig.~\ref{Fig-DeformedSphere}. The bound state is
essentially supported at the dip of the potential.
If the geometry is compact in the $X$ direction
(which is the case for de Sitter space or the CDL geometry,
but not for anti-de Sitter space or Minkowski space),
this mode is normalizable and
should be included in the complete set.
It can be shown that there is at most one bound state
in 3+1 dimensions~\cite{GarrigaMontes2, FSSY}\footnote{In higher
dimensions, it is possible to have
more than one bound states.}.
Normalization factor ${\cal N}$
in the thin-wall limit~\eqref{Euclida} is
\begin{align}
{\cal N}^{-2}&=\int^{\infty}_{-\infty}dX (H_Aa)^2(X)
=\int_{-\infty}^{X_0}dX \left({e^{X-X_0}\over
\cosh X_0}\right)^2+\int_{X_0}^{\infty}dX {1\over \cosh^2 X}\\
&={1\over 2\cosh^2 X_0}+1-\tanh X_0.
\end{align}
In the limit of $X_0\to -\infty$ (the limit of the small bubble),
we get a finite value ${\cal N}\to 1/\sqrt{2}$, which agrees
with the case for the pure de Sitter.

When the scalar field has mass, the additional term $+m^2a^2(X)$ lifts
the potential, and the dip in the potential becomes shallow.
The energy of the bound state (as long as exists) increases,
shifting the pole from $k_B=i$ in the massless case to
\begin{equation}
 k_B=i(1-\epsilon).
\end{equation}
If mass is larger than a certain value (which is of order $H_A$),
the bound state disappears.

We will consider the possibility that
mass of the field $\phi$ is different in the
true and the false vacuum,
\begin{equation}
m=\left\{ \begin{array}{ll}
m_A &\qquad (\mbox{False vacuum})\\
m_0 &\qquad (\mbox{True vacuum})\\
\end{array} \right. \ ,
\label{m}
 \end{equation}
 where $m_A$ and $m_0$ are assumed to be constant.
 The case of our interest is
 $m_A\ll H_A$ and $m_0\sim H_0\ll m_A$.
 In this case we can set $m_0=0$ in the analysis of
 the early-time behavior performed in this section,
 since such a tiny $m_0$ as compared to the natural scale
 (the Hubble parameter at the time in question) will not affect
the dynamics.

 It is not easy to calculate $\epsilon$ (i.e., the bound
 state energy) in general\footnote{The general expression in the
 thin-wall limit can be written in terms of
hypergeometric functions.},
 but when $m_A H_{A}^{-1}$ is small,
 as in the case of our interest, it can be evaluated
 by the first-order
perturbation theory. We take \eqref{uB} as the zeroth-order
wave function and $m^2 a^2(X)$ as perturbation Hamiltonian
in the one-dimensional  Schr\"{o}dinger-like equation
~\eqref{Schrodinger}.
The eigen-energy $E=(k_B^2+1)=1-(1-\epsilon)^2$
is zero at the zeroth order. The first-order
eigen-energy,
$E^{(1)}=2\epsilon$, is
\begin{equation}
 E^{(1)}=\int_{-\infty}^{\infty}dX
\left(\psi^{(0)}(X)\right)^2m^2a^2(X).
\label{E1}
\end{equation}
with the zeroth-order (massless) wave function,
\begin{equation}
 \psi^{(0)}(X)={\cal N} H_A a(X).
\end{equation}
Thus, for the case of $m_0^2=0$, we obtain
\begin{align}
 \epsilon &={{\cal N}^2\over 2}m_{A}^2 H_A^{-2}
\int_{X_0}^{\infty}dX{1\over \cosh^4X} \nonumber \\
&={{\cal N}^2\over 2}
\left({2\over 3}-{1\over 3}\left(2+\sech^2 X_0
\right)
\tanh X_0\right)m_{A}^2 H_A^{-2} \nonumber \\
&={2-(2+\sech^2 X_0)\tanh X_0\over 3\, (2+\sech^2 X_0-2\tanh X_0)}
m_{A}^2 H_A^{-2}.
\label{epsilonpert}
\end{align}

The reflection coefficient ${\cal R}(k)$ for
\eqref{Schrodinger} with $m_A=m_0=0$ in the thin-wall
limit \eqref{Euclidmetric} can be obtained exactly~\cite{FSSY},
\begin{equation}
 {\cal R}(k)={\gamma i e^{2ikX_0}\over (k+\gamma i)}{(k+i)\over (k-i)},
\label{Rkmassless}
\end{equation}
where $\gamma$ is given by
\begin{equation}
 \gamma={1\over 1+e^{-2X_0}},
\end{equation}
and takes a value $0<\gamma<1$. The $k=i$ pole
corresponds to the bound state.
The pole in the lower half plane ($k=-i\gamma$)
does not correspond to a physical mode for \eqref{Schrodinger},
since the wave function blows up at $X\to \pm \infty$ and is not
normalizable.

Using this complete set, the correlation function satisfying
\eqref{Greenfneq} can be expressed as
\begin{align}
\langle \chi(X, \theta)\chi (X', 0)\rangle
&=\int_0^{\infty} {dk\over 2\pi} \left(u^L_k(X)u^{L *}_k(X')
+u^R_k(X)u^{R *}_k(X')\right)G_k(\theta)\nonumber\\
&+u^B_k(X)u^{B *}_k(X') G_k(\theta),
\label{Euclideancorrelator0}
\end{align}
where $G_k(\theta)$ is the Green's function
on $S^3$ for a field with the effective mass $(k^2+1)$,
\begin{equation}
 \left[-\nabla_S^2+(k^2+1)\right]G_k(\theta)={\delta(\theta)
  \over 4\pi \sin^2\theta}.
  \label{Gkthetaeq}
\end{equation}
$G_k(\theta)$ is a function of
the geodesic distance $\theta$ on
$S^3$, and given by
\begin{equation}
 G_k(\theta)={\sinh k(\pi-\theta)\over 4\pi\sinh k\pi \sin\theta}.
\label{Gktheta}
\end{equation}
We can see that $G_k(\theta)$ is the correct Green's function on $S^3$
by noting the following facts. $G_k(\theta)$ satisfies the Laplace
equation, and is regular everywhere
on $S^3$ except for the source $\theta=0$ with the correct singularity
$G_k(\theta)\sim 1/(4\pi\theta)$ at
$\theta=0$.

There is a subtlety when
the mass of the scalar field $\phi$ is exactly zero.
We have a bound state with $k=i$, and
the effective mass on $S^3$
becomes zero, $k^2+1=0$. But there does not exist such a
 Green's function on a compact space, since the flux
cannot extend to infinity and the Gauss law cannot be satisfied.
(In other words, the differential operator in the
equation of motion cannot be inverted in this
case\footnote{In non-compact spaces, this does not cause a
problem, since the zero mode is measure zero.},
since the equation is unchanged under the constant shift
of $\phi$). See \cite{FSSY}
for how to deal with this case. In this paper, we will not have
this problem, since we will always consider the massive field.

\begin{figure}[!htb]
\center
\includegraphics [scale=.3]
 {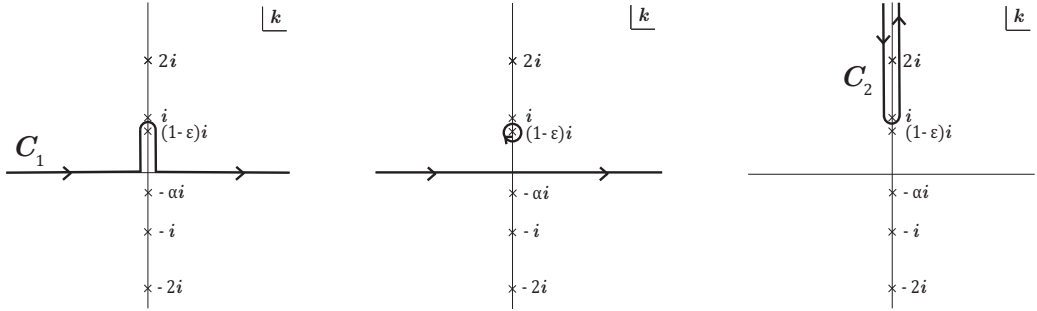}
 \caption{Left panel: The definition of the contour ${\cal C}_1$ for
 the $k$-integration. Middle panel: A contour equivalent to
 ${\cal C}_1$.
 The residue at the $k=(1-\epsilon)i$ pole is
equal to the bound state contribution (the second
 line) in \eqref{Euclideancorrelator0}. Right panel: The contour
 ${\cal C}_2$ used at the end of Section 4 to show the
 finiteness of $\langle \phi^2\rangle$ in the early-time
($\eta, \eta'\to -\infty$) limit. By deforming
 ${\cal C}_1$ to ${\cal C}_2$, the $k$-integral is expressed
 as a sum over the residues at $k=in$ ($n=1,2,\ldots$).}
 \label{Fig-contour}
\end{figure}

The correlation function in the $X\to -\infty$ limit
 (i.e., on the true vacuum side)
when $m_0=0$, can be written as
\begin{equation}
\langle \chi(X, \theta)\chi (X', 0)\rangle
=\int_{{\cal C}_1} {dk\over 8\pi^2} \left(e^{ik(X-X')}+{\cal R}(k)e^{-ik(X+X')}\right)
{\sinh k(\pi-\theta)\over \sinh k\pi \sin\theta}.
\label{Euclideancorrelator}
\end{equation}
If we take the thin-wall limit,
\eqref{Euclideancorrelator} is valid
throughout the $X<X_0$ region.
The integration contour ${\cal C}_1$ is defined
in  Fig.~\ref{Fig-contour}. This contour is basically along the real axis
but is deformed near the origin so that it goes above
the bound state pole.
This deformation amounts to adding a discrete mode due to
the bound state.

Note that the normalization factor $1/\sinh k\pi$ for the $S^3$ Green's
function $G_k(\theta)$ introduces poles at integer multiple
of $k=i$. Thus, if the mass is exactly zero,
the pole at $k=i$ becomes a double pole. As mentioned above,
we will always consider the massive case, so
the contour ${\cal C}_1$ passes between
$k=i(1-\epsilon)$ and $k=i$. We can take the massless limit
starting from this expression, if necessary.

\subsection{Analytic continuation to Lorentzian}

We perform the analytic continuation \eqref{analyticcont2}
on \eqref{Euclideancorrelator}, and obtain the correlation function
in the FLRW universe\footnote{In \eqref{FRWcorrelator0},
the integration contour for the first term has been deformed
from ${\cal C}_1$ to the real $k$ axis. Since this
term does not have a bound state pole, one can
freely make this deformation.},
\begin{align}
\langle \chi(\eta, R)\chi (\eta', 0)\rangle&
=\int_{-\infty}^{\infty} {dk\over 8\pi^2}
e^{ik(\eta-\eta')}
{\sinh k(\pi-i R)\over i \sinh k\pi \sinh R}\nonumber\\
&+\int_{{\cal C}_1} {dk\over 8\pi^2} {\cal R}(k)e^{-ik(\eta+\eta')}
{\sin k R\over \sinh k\pi \sinh R}.
\label{FRWcorrelator0}
\end{align}
This is valid in the early-time limit $\eta, \eta'\to -\infty$
(i.e., the curvature dominated era), and
can be regarded as the
initial condition for the correlator in the FLRW universe.
The correlator at later times will be obtained in Section~5.

A subtle prescription for analytic continuation has been used
to obtain the second term in \eqref{FRWcorrelator0}. If we naively
made the substitution \eqref{analyticcont2} in the second term in
\eqref{Euclideancorrelator}, we would have gotten
\begin{equation}
e^{-ik(X+X')}\, {\sinh k(\pi-\theta)\over \sinh k\pi}
\to e^{-ik(\eta+\eta')}\, {\left( e^{2k\pi-ikR}-e^{ikR}\right)
\over 2 \sinh k\pi},
\label{naive}
\end{equation}
but the first term in the numerator
on the right hand side of \eqref{naive}
diverges for $k\to \infty$, so the $k$ integral along the
real axis does not converge. To get a convergent integral,
we should make a replacement
\begin{equation}
\sinh k(\pi-\theta)\to e^{-k\pi}\sinh \theta
\label{replacement}
\end{equation}
in
\eqref{Euclideancorrelator}
before the analytic continuation.
This is allowed, since the replacement~\eqref{replacement}
does not change the result of the
integration in
\eqref{Euclideancorrelator}.
The integral can be evaluated
by deforming the contour and
summing up the residues of the poles at $k=ni$ with
$n=1, 2, \ldots$.
As a result, multiplication by $e^{2k\pi}$ amounts
to multiplication of each residue by unity, so the answer does
not change. By performing analytic continuation after
the replacement \eqref{replacement}, we obtain the
second term in \eqref{FRWcorrelator0}. On the other hand,
for the first term
in \eqref{Euclideancorrelator}, analytic continuation
should be done using the original expression.  The
$k$-integral converges with this integrand, but does not
converge if we make the replacement \eqref{replacement}.

In fact, evaluating the integral \eqref{Euclideancorrelator}
as sum over the poles at $k=ni$ ($n=1, 2, \ldots$) is equivalent
to expanding the correlator into spherical harmonics on
$S^3$. Total angular momentum $L$ on $S^3$ is related to
the position of the pole by $L=n-1$. For the
calculation of the correlator starting from this representation
of discrete sum, and using Watson-Sommerfeld transformation to
convert it into an integral, see e.g., \cite{Turok1, Turok2}.

The correlator \eqref{FRWcorrelator0} is a
function of the geodesic distance $R$ on $H^3$.
In other words, one point is set at the origin
and the other point is at a radial distance $R$
from the origin. When two points are at general positions
on $H^3$, we can rewrite the correlator using the relation
\begin{equation}
 \cosh R=\cosh R_1 \cosh R_2 -\sinh R_1\sinh R_2 \cos\psi.
\end{equation}
where $R_1$ and $R_2$ are the radial coordinates of the
two points, and  $\psi$ is the angle on $S^2$ between them.

The correlator can also be expressed as a sum over
products of harmonics $f^{(\ell, m)}_{k}(R, \Omega_2)$
at the two points. Harmonics
on $H^3$ are the eigenfunctions of Laplacian,
\begin{equation}
 \nabla_H^2 f^{(\ell, m)}_{k}(R, \Omega_2)
  =-(k^2+1)f^{(\ell, m)}_{k}(R,\Omega_2).
\label{LapfkR}
\end{equation}
The modes with real $k$ are
normalizable on $H^3$,
\begin{equation}
\label{norm}
 \int_0^{\infty} dR \sinh^2 R\int d\Omega_2
  f^{(\ell, m)}_k(R, \Omega_2)
  f^{(\ell', m')*}_{k'}(R, \Omega_2)=
  \delta(k-k')\delta_{\ell, \ell'}\delta_{m, m'}.
\end{equation}

The $\ell=0$ mode, which is homogeneous on $S^2$,
is a function of only the radial coordinate,
\begin{equation}
 f^{(\ell=0)}_{k}(R)=
  \frac{1}{\sqrt{2}\pi}{\sin k R\over \sinh R}\ .
\label{fkR}
\end{equation}
When
one point is at the origin $R_1=0$ or $R_2=0$,
only the $\ell=0$ mode contributes to the two-point functions (as
is familiar in the harmonic expansion in quantum mechanics in flat
space). The factor $\sinh R$ in the denominator of \eqref{fkR}
compensates for the exponential growth of the volume for large $R$ on
the hyperboloid. For the explicit form of the harmonics with
$\ell\neq 0$, see Appendix~\ref{sec-modes} and
\cite{SasakiTanakaYamamoto1,SasakiTanakaYamamoto2,
SasakiTanakaYamamoto3}.

Note that a constant function on $H^3$ is not normalizable.
To express non-normalizable functions, which decay
more slowly than $e^{-R}$ as $R\to \infty$ (with
arbitrary dependence on $S^2$),
we need the modes with imaginary $k$.
As indicated by the integration contour ${\cal C}_1$,
the correlator \eqref{FRWcorrelator0} contains a
discrete non-normalizable mode on $H^3$, which is called the
``supercurvature mode''~\cite{SasakiTanakaYamamoto1,SasakiTanakaYamamoto2,
SasakiTanakaYamamoto3}.
As we have seen, a bound state in the Euclidean problem
is in one-to-one correspondence with a
supercurvature mode on $H^3$.

The first term  in \eqref{FRWcorrelator0}, which is
a function of $\eta-\eta'$,
is nothing but the correlator in global Minkowski space
written in the open slicing. This can be seen
by rewriting the massless correlator in terms of Minkowski
coordinates $\hat{t}$, $\hat{r}$,
and performing the coordinate transformation \eqref{hatt2},
 \begin{align}
  \langle\phi(\eta, R)\phi(\eta', 0)\rangle&=
{ 1\over \left(-(\hat{t}-\hat{t'})^2+\hat{r}^2\right)}\nonumber\\
  &={1\over 2e^{\eta+\eta'}\left(\cosh R-\cosh (\eta-\eta')\right)},
\label{flat}
 \end{align}
 where we have put one point at the origin $\hat{r}'=0$.
 Carrying out the $k$ integral for the first term
 in \eqref{FRWcorrelator0} as a sum over the residues from the
 poles, we recover the r.h.s. of \eqref{flat}.

 The second term, which is a function on $\eta+\eta'$ and
 depends on the reflection coefficient
${\cal R}(k)$, carries the information about the ancestor vacuum.
This term is finite in the coincident-point limit
$R\to 0$, $\eta-\eta'\to 0$.

\section{Energy-momentum tensor}
\label{sec-EMT}

We  now compute the contribution of a quantum field to
the energy-momentum tensor, by taking the coincident-point
limit of the two-point function obtained in the
previous section.

\subsection{General remarks}
The energy-momentum tensor for a minimally coupled scalar field is
 \begin{equation}
  T_{\mu\nu}=\partial_\mu\phi\partial_\nu\phi
   -{1\over 2}g_{\mu\nu}\left(
\partial^\rho\phi\partial_\rho\phi+m^2\phi^2\right).
\label{Tmunu1}
 \end{equation}
The energy density $\rho=-\langle T^\eta_{~\eta}\rangle$ and
pressure $p=\langle T^i_{~i}\rangle$, where $i$'s
are not summed over,
are obtained by taking the vacuum expectation
value of (\ref{Tmunu1}):
\beqa
\rho&=& \frac{1}{a^2} \left\langle \frac{1}{2} \phi'^2
+\frac{1}{2} (\nabla\phi)^2 +\frac{1}{2} m^2a^2 \phi^2
\right\rangle \ , \label{rho1}\\
p&=&\frac{1}{a^2} \left\langle \frac{1}{2} \phi'^2
-\frac{1}{6} (\nabla\phi)^2
-\frac{1}{2} m^2a^2 \phi^2 \right\rangle \ .
\label{p1}
\eeqa
They can be computed from
the two-point function in the the coincident-point limit.

The two-point function
\eqref{FRWcorrelator0} consists of two terms.
The first term is independent of the ancestor vacuum.
In the early-time limit, it coincides with the correlator
in the global Minkowski space. We will touch on this term only
briefly in this paper, since the analysis of this term
is essentially contained in the previous
work \cite{AIS,AI}.

The second term in \eqref{FRWcorrelator0} depends on the
properties of the ancestor vacuum. The supercurvature
mode is involved in this term.
We will mostly focus on this second term, which is 
larger than the first term in the
cases of interest, as we will see later. The value of the second term
 is found to be finite
in the coincident-point limit, and thus is independent of
the renormalization prescription for the UV divergence.

The wave functions in the ancestor vacuum are of order
$H_A$, the natural mass scale in de Sitter space.
As we will see in
 detail in Section~\ref{sec-evolution},
the continuous modes have the time dependence $e^{-\eta}$,
and they decreases to order $H_I$, which is the
 natural magnitude of fluctuations in slow-roll
 inflation with the Hubble parameter $H_I$,  by the end of
 the curvature domination~\cite{FKMS1}.
The supercurvature mode has the time dependence $e^{-\epsilon \eta}$,
and decays more slowly than the continuous modes. Furthermore,
its contribution to $\langle \phi^2 \rangle$ is enhanced by an
extra factor $\epsilon^{-1}\sim H_A^2/m_A^2$,
as we will see shortly.

The continuous modes are different from the modes in the
flat FLRW cosmology, in the sense that the infrared modes
are cut off by the spatial curvature.
We can see this e.g., in \eqref{fkR} where the mode
with real $k$ decays
exponentially for $R\gtrsim 1$.

\subsection{Mass term in the limit of small mass}

Let us study the mass term in the limit of small mass
($\epsilon$), which will be the most
dominant term in the energy density of the scalar field.

The correlation function in the early
time limit $\eta, \eta'\to -\infty$ is
\begin{align}
\langle \chi(\eta, R)\chi (\eta', 0)\rangle&
=\int_{-\infty}^{\infty} {dk\over 8\pi^2}
e^{ik(\eta-\eta')}
{\sinh k(\pi-i R)\over i \sinh k\pi \sinh R}\nonumber\\
 &+\int_{{\cal C}_1} {dk\over 8\pi^2}
 {\cal R}(k)e^{-ik(\eta+\eta'+2\tilde{\eta}_1)}
{\sin k R\over \sinh k\pi \sinh R},
\label{FRWcorrelator}
\end{align}
where we have made
the shift $\eta\to \eta+\tilde{\eta}_1$
with $\tilde{\eta}_1$
defined in \eqref{tildeeta} (similarly for $\eta'$)
in \eqref{FRWcorrelator0},
to account for the difference in the definition of $\eta$
mentioned in Section 2.

The contribution from the $k=i(1-\epsilon)$ pole is
\begin{equation}
 \langle \chi(\eta, R)\chi (\eta', 0)\rangle^{\rm (s.c.m.)}
={(-2\pi i)\over 8\pi^2}\cdot \mbox{Res}(i(1-\epsilon))e^{(1-\epsilon)(\eta+\eta'
+2\tilde{\eta}_1)}{1\over \sin \epsilon\pi}
{\sinh(1-\epsilon)R\over \sinh R},
\label{chichiNNM}
\end{equation}
where $\mbox{Res}(i(1-\epsilon))$ denotes the residue
of $R(k)$ at $k=i(1-\epsilon)$. This gives the contribution from the supercurvature mode on $H^3$. Dividing
\eqref{chichiNNM} by $a(\eta)a(\eta')$, and taking the
coincident-point limit $\eta\to \eta'$, $R\to 0$, we
obtain the expectation value
$\langle \phi^2(\eta)\rangle^{\rm (s.c.m.)}$.

Let us explicitly
evaluate $\langle \phi^2(\eta)\rangle^{\rm (s.c.m.)}$
in the $\epsilon\to 0$ limit. In this limit, one can use
the residue of $R(k)$ for $\epsilon=0$,  given by
\eqref{Rkmassless}, and obtain
\begin{equation}
 \langle \phi^2 (\eta)\rangle^{\rm (s.c.m.)}
  ={1\over 4\pi^2\epsilon}A(X_0)
e^{-2\epsilon(\eta+\tilde{\eta}_1)}H_{A}^2\ ,
\label{phiphiNNM}
\end{equation}
where $A(X_0)$  depends on $X_0$
(i.e., the size of the bubble),
\begin{equation}
 A(X_0)={(1+e^{2X_0})^2\over 2(1+2e^{2X_0})}.
\label{AX0}
\end{equation}
We have kept $\epsilon$ in the denominator and
in the time-dependent function
$e^{-2\epsilon(\eta+\tilde{\eta}_1)}$, but have replaced
the term
$e^{2(1-\epsilon)X_0}$ by $e^{2X_0}$, since this
difference will not have a large effect.
Using the value of $\epsilon$ obtained by the
first-order perturbation theory in \eqref{epsilonpert},
the expectation value \eqref{phiphiNNM} becomes
\begin{equation}
 \langle \phi^2 (\eta)\rangle^{\rm (s.c.m.)}
={1\over 2\pi^2}\cdot {3\, e^{-2\epsilon(\eta+\tilde{\eta}_1)}\over 2-(2+\sech^2 X_0)\tanh
  X_0}\cdot {H_{A}^4\over m_{A}^2}.
\label{phiphi2}
\end{equation}

In the limit of the small bubble $X_0\to -\infty$,
(and $\epsilon\to 0$), \eqref{phiphi2} approaches ${3\over 8\pi^2}{H_A^4 \over m_A^{2}}$.
This agrees with the well-known result in
pure de Sitter space obtained by the standard techniques
(see e.g., \cite{BirrellDavies, Linde}). For the convenience of the
 reader, we summarize the calculation of $\langle \phi^2\rangle$
 for the pure de Sitter case
 in Appendix~\ref{sec-puredS}.

One may worry that $\langle \phi^2(\eta)\rangle^{\rm (s.c.m.)}$
given in \eqref{phiphiNNM} or \eqref{phiphi2} diverges
in the early-time limit $\eta\to -\infty$. In fact, the full
expression $\langle \phi^2(\eta)\rangle$ also including the continuous
modes is regular, as can be shown by deforming the contour for
the $k$-integration: At early times,
the $e^{-ik(\eta+\eta')}$ factor
in $\langle \chi(\eta)\chi(\eta')\rangle$ gives a
converging factor for ${\rm Im}(k)>0$, allowing us to deform
the contour ${\cal C}_1$ into ${\cal C}_2$ defined
on the right panel in Fig.~\ref{Fig-contour}. Then
the $k$-integral is represented as a sum over the residues
at the poles at $k=i n$ $(n=1,2,\ldots)$. The time-dependence
of each residue in $\langle \chi(\eta)\chi(\eta')\rangle$
is given by $e^{-ik(\eta+\eta')}= e^{n(\eta+\eta')}$.
The leading term in the early-time limit comes from $n=1$,
giving
$\langle \phi^2(\eta)\rangle=
\langle \chi^2(\eta)\rangle/a^2(\eta)\sim {\rm const.}$

\section{Evolution of vacuum energy}
\label{sec-evolution}

Having obtained
the vacuum expectation value of the energy-momentum
tensor in the early-time limit in the FLRW universe,
we now study its time evolution. First, in Section~\ref{subsec:scalefactor},
we determine the scale factor for the background open universe.
Then, in Section~\ref{subsec:wavefunction},
we obtain the wave function by solving
the equation of motion for the scalar field,
and in Section~\ref{subsec:energydensity}, we study the evolution of the vacuum
expectation value of the energy-momentum tensor.
Although our analysis is general,
we will mostly focus on the supercurvature mode.
Finally, in Section~\ref{subsec:vaceneDE}, we consider
the possibility that the vacuum energy gives the present
dark energy.

\subsection{The scale factor}
\label{subsec:scalefactor}

The scale factor for an open universe can be found by
solving the Friedmann equation,
\begin{equation}
{(a')^2\over a^4}={1\over 3M_P^2}\rho+{1\over a^2},
\label{Friedmann2}
\end{equation}
with the energy density $\rho$ appropriate for each era.

The universe at early times is curvature dominated (CD).
As we mentioned in section~\ref{subsec-FLRW},
when the vacuum energy of inflation dominates
over the curvature contribution,
the slow-roll inflation occurs.
The scale factor in the CD and Inflation eras is given in (\ref{aInf}).
After reheating, the universe
becomes radiation dominated (RD). For simplicity, we assume
the reheating occurs instantaneously. In fact, the reheating process
takes place for producing the particles, mainly of sub-horizon modes,
which the inflation field is coupled to. Thus the detailed dynamics
of particle production will be
irrelevant to our main concern on the energy
density  given by the supercurvature mode.
 Then, after the matter-radiation equality, the
universe becomes matter dominated (MD).

The scale factors in each era are obtained by solving (\ref{Friedmann2}) as
\begin{equation}
a(\eta)=\left\{ \begin{array}{lll}
a_{\rm CDInf}(\eta)=-\frac{1}{H_I \sinh (\eta)} &(-\infty<\eta<\eta_1<0) &(\mbox{CD-Inflation})  \\
a_{\rm RD}(\eta)=\alpha \sinh(\eta) &(0<\eta_2<\eta<\eta_3)&(\mbox{RD}) \\
a_{\rm MD}(\eta)=\beta\sinh^2(\eta/2)  &(\eta_4<\eta<\eta_0) &(\mbox{MD})
\end{array} \right. 
\label{aInfRDMD}
\end{equation}
with $\rho=3M_p^2 \alpha^2/a^4$ for the RD era
and $\rho=3M_p^2 \beta/a^3$ for the MD era.\footnote{
The solution for the scale factor when $\rho$ contains
both radiation and matter
is described in Appendix~\ref{sec-radmatt}.}.
We will connect them by requiring
$a(\eta)$ and $a'(\eta)$ are continuous across each era.
Here, we closely follow the convention of
ref.~\cite{AIS}\footnote{$\beta$ here corresponds
 to $4\beta$ in \cite{AIS}.}.
Namely, we introduce the different
 shift of $\eta$ in each era, to keep the expression
 of $a(\eta)$ simple.
The conformal time at present is $\eta=\eta_0$.
 We ignore the present
cosmological constant (dark energy) and assume the MD
era continues until now.

The continuity conditions for $a$ and $a'$ give the following relations
among the parameters in (\ref{aInfRDMD}):
\beqa
\eta_1&=&-\eta_2 \ , \label{eta1eta2}  \\
\sinh^2(\eta_1) &=& \frac{1}{\alpha H_I} \ , \label{eta1alphaHI} \\
\eta_3&=&\frac{\eta_4}{2}  \ , \label{eta3eta4}  \\
\sinh(\eta_3) &=& \frac{\alpha}{\beta} \ . \label{eta3alphabeta}
\eeqa
The Hubble parameter in each era is calculated from (\ref{aInfRDMD}) as
\beq
H=\frac{a'}{a^2} =\left\{ \begin{array}{ll}
H_I \cosh(\eta) &(\mbox{CD-Inflation})  \\
\frac{\cosh(\eta)}{\alpha\sinh^2(\eta)} &(\mbox{RD}) \\
\frac{\cosh(\eta/2)}{\beta\sinh^3(\eta/2)} &(\mbox{MD})
\end{array} \right. \ .
\label{HInfRDMD}
\eeq
Comparing the Hubble parameters at the beginning and the
end of each era, we obtain a relation
\beq
\sinh^2(\eta_2) \sinh(\eta_3) \frac{\cosh(\eta_0/2)}{\sinh^3(\eta_0/2)}
= \frac{H_0}{H_I} \ ,
\eeq
where $H_0$ is the present Hubble parameter.
This can also be obtained by multiplying (\ref{eta1alphaHI}), (\ref{eta3alphabeta}),
and (\ref{HInfRDMD}) at present time.
From observations~\cite{PlanckCosm}, we know
\beqa
&&H_0 \simeq 67 {\rm km}~{\rm s}^{-1}{\rm Mpc}^{-1} \simeq 1.4\times10^{-30}{\rm meV} \ ,
\label{H0obsdata} \\
&&H_I \lesssim 3.6\times 10^{-5} M_P \simeq 8.8\times 10^{25}{\rm meV} \ ,
\label{HIupperbound}
\eeqa
where meV=$10^{-3}$eV, and
$M_P = (8 \pi G_N)^{-1/2} \simeq 2.4\times 10^{30} {\rm meV}$ is
the (reduced) Planck scale.

The ratio of the curvature radius $R_c=a$ (note that curvature
radius is unity in the comoving coordinate) to the Hubble radius
$H^{-1}=a^2/a'$ is
\beq
r_c \equiv {R_c\over H^{-1}}={a'\over a}=
\left\{ \begin{array}{ll}
-\coth(\eta) &(\mbox{CD-Inflation})  \\
\coth(\eta)&(\mbox{RD}) \\
\coth(\eta/2)&(\mbox{MD})
\end{array} \right. \ .
\label{NcInfRDMD}
\eeq
This ratio grows during inflation, and decreases during
the RD and the MD eras. $r_c$ is a huge number at the end of inflation,
and remains large until now due to the observational bound
at present time~\cite{PlanckCosm},
\begin{equation}
r_{c0} \gtrsim 10\quad (\mbox{now}).
\label{Rcbound}
\end{equation}
The redshift parameter at the
matter-radiation equality is
\beq
z_{\rm eq} \simeq 3.4 \times 10^3 \ ,
\label{zeqobs}
\eeq
and thus we have
\beq
1+z_{\rm eq} = \frac{a_0}{a_4}=\frac{\sinh^2(\eta_0/2)}{\sinh^2(\eta_4/2)} \ .
\label{zeqdef}
\eeq

Using the above relations, we now estimate the values of the parameters
 $\eta_1$, $\eta_2$, $\eta_3$, $\eta_4$,
$\eta_0$, $\alpha$ and $\beta$ in (\ref{aInfRDMD}).
From (\ref{NcInfRDMD}) at present and (\ref{zeqdef}),
the values of $\eta_0$ and $\eta_4$ are constrained as
\beqa
\eta_0 &\simeq& 2r_{c0}^{-1} \lesssim 0.2 \label{eta0Nc0} \ , \\
\frac{\eta_4}{\eta_0} &\simeq& z_{\rm eq} ^{-1/2} \simeq
1.7 \times 10^{-2} \ ,
\eeqa
where (\ref{Rcbound}) and (\ref{zeqobs})
are used.
Then, from (\ref{eta3alphabeta}), and (\ref{HInfRDMD}) at present,
the values of $\alpha$ and $\beta$ are obtained,
\beqa
\frac{\alpha}{\beta} &\simeq& \frac{\eta_4}{2} \simeq
r_{c0}^{-1}  z_{\rm eq} ^{-1/2} \lesssim  2 \times10^{-3} \ ,
\label{albeest}\\
\beta H_0 &\simeq & \left(\frac{\eta_0}{2}\right)^{-3}
\simeq r_{c0}^3 \gtrsim 1 \times 10^3 \ .
\label{betaest}
\eeqa
Finally, with (\ref{eta1alphaHI}), the value of $\eta_2$
is estimated as
\beq
\eta_2 \simeq
\left(\frac{\beta}{\alpha} \frac{1}{\beta H_0}\frac{H_0}{H_I}\right)^{1/2}
\simeq r_{c0}^{-1} z_{\rm eq}^{1/4} \left(\frac{H_0}{H_I}\right)^{1/2}
\sim 10^{-28} \ ,
\label{eta2est}
\eeq
where the lower bound for $r_{c0}$ and the upper bound for $H_I$
are applied in the last equality.

Slow-roll inflation starts when the inflaton potential
starts to dominate over
the curvature term, i.e., at $\eta\sim -1$.
The e-folds $N_e$ of the slow-roll
inflation is
\beq
N_e=\log \left(\frac{a_{\rm CDInf}(\eta_1)}{a_{\rm CDInf}(-1)}\right)
\sim 64.6 \ ,
\label{Neest}
\eeq
where we have used (\ref{eta2est}) for the value of $\eta_1=-\eta_2$.
If we take larger values for $r_{c0}$ (the ratio between the
curvature radius and Hubble radius at present),
$N_e$ becomes larger. Since the value (\ref{Neest}) roughly
coincides with the value derived from an anthropic bound for
the spatial curvature~\cite{FKMS1},
it would be reasonable to adopt these minimal values for
$r_{c0}$ and $N_e$.

\subsection{Wave functions}
\label{subsec:wavefunction}

\subsubsection{Equation of motion}
We now consider a minimally coupled scalar field $\phi$
in the open FLRW universe. The equation of motion
for $\chi$, which is defined as $\phi=\chi/a$, is
\begin{equation}
 \left[-\partial_\eta^2+{a''\over a}
+\nabla_H^2-m_0^2a^2\right]\chi(\eta, {\cal H})=0 \ ,
\label{Schrodinger1}
\end{equation}
where ${\cal H}$ and $\nabla_H^2$
denote the coordinates and Laplacian, respectively, on $H^3$ .
The mass term in \eqref{Schrodinger1} is the
mass $m_0$ after tunneling.

We expand the field into harmonics on $H^3$,
defined in \eqref{LapfkR},
\beq
\chi(\eta, {\cal H}) =
\int_0^{\infty} dk \sum_{\ell=0}^{\infty}\sum_{m=-\ell}^{\ell}
v_k(\eta) f^{(\ell,m)}_k({\cal H})+
\sum_{\ell=0}^{\infty}\sum_{m=-\ell}^{\ell}
v_*(\eta) f^{(\ell,m)}_{k_B}({\cal H})\ ,
\eeq
where the first term is the contribution from the continuous
modes and the second term is from the
supercurvature mode. $v_k(\eta)$ and $v_*(\eta)$ are
the time-dependent part of the wave functions
for the continuous modes and the
supercurvature mode.
(The corresponding parts in $\phi$ will be called
$\varphi_k=v_k/a$ and
$\varphi_*=v_*/a$.)
The harmonics are the eigenfunctions
of Laplacian with eigenvalues $\nabla_H^2\to -(k^2+1)$.
For the supercurvature mode, we have
$k=k_B\equiv i(1-\epsilon)$, where $\epsilon$ depends
on the mass $m_A$ in the ancestor vacuum, and is of
order $m_A^2 H_A^{-2}$ in the small $\epsilon$ limit,
as explained in Section~3.1.

Each mode on $H^3$ is independent at the linearized level.
Thus, we will concentrate on the supercurvature mode.
The wave function $v_*(\eta)$ satisfies
\begin{equation}
 \left[-\partial_\eta^2+{a''\over a}-1+(1-\epsilon)^2
-m_0^2a^2\right]v_*(\eta)=0 \ .
\label{eqetascm}
\end{equation}
This is of the form of the time-independent
Schr\"{o}dinger equation,
like the Euclidean equation of motion studied in Section~3.

From
\eqref{Friedmann} and \eqref{conservation}, we find
\begin{equation}
{a''\over a}-1={a^2\over 6M_p^2}(\rho-3p)={a^2\over 6M_p^2}(1-3w)\rho.
\label{abound}
\end{equation}
Thus, as long as $w\le 1/3$, which is the case for
the CD era, inflation, the RD and the MD eras,
the ``potential'' in~(\ref{eqetascm}) is
non-negative, $a''/a-1\ge 0$. The ``eigen-energy'' of the
supercurvature mode,
$k^2=-(1-\epsilon)^2$, is negative and is always below
the potential. Thus the massless wave function is not oscillating.

In Sections~\ref{subsec:scalar} and \ref{subsec:epsilon0}
below, we will study Eq.~(\ref{eqetascm}) by making an
approximation, retaining the relevant two terms from the
following three terms,
$a''/a$, $1-(1-\epsilon)^2=2\epsilon-\epsilon^2$,
and $m_0^2a^2$.
Since $a''/a>1$ is satisfied in all the eras of interest
(as shown in (\ref{abound}) and also in (\ref{PotentialInfRDMD})
below), $a''/a>2\epsilon-\epsilon^2$ is always satisfied.
Thus, there are the following three cases to be considered:
\begin{enumerate}
\item[(i)] $m_{0}^2a^2<2\epsilon-\epsilon^2<a''/a$
\item[(ii)] $2\epsilon-\epsilon^2<m_{0}^2a^2<a''/a$
\item[(iii)] $2\epsilon-\epsilon^2<a''/a<m_{0}^2a^2$
\end{enumerate}
As time passes and the scale factor $a$ increases,
the universe undergoes the periods (i), (ii) and (iii)
in turn.

Before starting the analysis, let us recall the well-known
qualitative
properties
of Eq.~(\ref{eqetascm}).
In the periods (i) and (ii), the term $a''/a$
is the largest in comparison with
$2\epsilon-\epsilon^2$ and $m_0^2a^2$.
If we keep only this term in \eqref{eqetascm},
the general solution is
\beq
\varphi_*=\frac{v_*}{a}=c_1 +c_2 \int \frac{d\eta}{a^2} \ ,
\eeq
where $c_1$ and $c_2$ are integration constants.
The first term shows {\it frozen} behavior of the wave function
while the second term describes a decaying mode.
Thus, for general cases with $c_1 \ne 0$, after a sufficiently
long duration, the wave function becomes frozen.
On the other hand, in the period (iii),
the term $m_0^2a^2$ is the most relevant.
If we only keep this term in (\ref{eqetascm}),
the solution is
\beq
\varphi_*=\frac{v_*}{a}\simeq \frac{1}{a^{3/2}} e^{\pm im_0 \int d\eta~a}
=\frac{1}{a^{3/2}} e^{\pm im_0t} \ .
\eeq
The wave function is {\it oscillating} and {\it decreasing}.
Owing to the frozen behavior of the wave function in the
periods (i) and (ii),
vacuum fluctuations generated in the ancestor vacuum
remain until late times,
and may provide the present dark energy.
At later times, in the period (iii),
the mass term will be relevant, and the vacuum energy
will diminish.

In the following, we will study Eq.~(\ref{eqetascm}) in more detail
by taking into account not only the first but also
the second largest of the three terms,
$a''/a$, $2\epsilon-\epsilon^2$, and $m_0^2a^2$.

\subsubsection{The massless approximation}
\label{subsec:scalar}

In the period (i),
we approximate (\ref{eqetascm}) by neglecting the mass term.
The scale factor (\ref{aInfRDMD}) gives rise to
the following potential in the Schr\"{o}dinger-like equation
(\ref{eqetascm}) in each era,
\begin{equation}
 {a''\over a}-1=\left\{ \begin{array}{ll}
{2\over \sinh^2 (\eta)} &(\mbox{CD-Inflation})  \\
0&(\mbox{RD}) \\
{1\over 2 \sinh^2 (\eta/2)} &(\mbox{MD})
\end{array} \right. \ .
\label{PotentialInfRDMD}
 \end{equation}
Then, the solutions are given by
\begin{equation}
 v_*=\left\{ \begin{array}{ll}
 \left(\pm (1-\epsilon)-{1\over \tanh(\eta)}\right)
e^{\pm(1-\epsilon)\eta} &(\mbox{CD-Inflation})  \\
e^{\pm(1-\epsilon)\eta} &(\mbox{RD}) \\
 \left(\pm 2(1-\epsilon)-{1\over \tanh(\eta/2)}\right)
e^{\pm(1-\epsilon)\eta} &(\mbox{MD})
\end{array} \right. \ ,
\label{chiscmInfRDMD}
 \end{equation}
 where the normalization of the wavefunction will be
 considered later.
 These solutions can also be obtained simply by the replacement $k\to i(1-\epsilon)$
from the wave function solutions of the continuous modes,
\begin{equation}
 v_k=\left\{ \begin{array}{ll}
 \left(\mp ik-{1\over \tanh(\eta)}\right)
e^{\mp ik\eta} &(\mbox{CD-Inflation})  \\
e^{\mp ik\eta}&(\mbox{RD}) \\
\left(\mp 2ik-{1\over \tanh(\eta/2)}\right)
e^{\mp ik\eta} &(\mbox{MD})
\end{array} \right. \ .
\label{chiconInfRDMD}
\end{equation}

We require the solution (\ref{chiscmInfRDMD}) to
match smoothly onto
(\ref{phiphi2}) found from the CDL geometry.
This selects the solution
with $e^{+(1-\epsilon)\eta}$ in the CD-Inflation era
in (\ref{chiscmInfRDMD}).
The wave functions in the RD and the MD eras are obtained by requiring
continuity conditions for $v_*$ and $v'_*$
across each era.
We thus have the following {\it normalized} wave functions:
\begin{equation}
 v_*=\left\{ \begin{array}{ll}
 N_* \left(1-\epsilon-{1\over \tanh(\eta)}\right)
e^{(1-\epsilon)\eta} &(\mbox{CD-Inflation})  \\
A~e^{(1-\epsilon)\eta} +B~e^{-(1-\epsilon)\eta}&(\mbox{RD}) \\
C \left(2(1-\epsilon)-{1\over \tanh(\eta/2)}\right)
e^{(1-\epsilon)\eta}
+D \left(-2(1-\epsilon)-{1\over \tanh(\eta/2)}\right)
e^{-(1-\epsilon)\eta}
&(\mbox{MD})
\end{array} \right. \ .
\label{chiscmmatch}
\end{equation}
To obtain the two-point function
$\langle\phi\phi\rangle^{\rm s.c.m.}$,
we replace the factor $e^{-\epsilon\eta}$ in the early
time expression \eqref{phiphi2}, by the solution $v_*/a$
at late times with a suitable normalization explained below.
We determine the coefficients $A$ to $D$ and obtain
the wave functions $v_*$
in the RD and MD eras in Appendix~\ref{sec-matwf}.

We now summarize the wave functions $\varphi_*=v_*/a$.
In the CD and Inflation eras, the wave function
is given by (\ref{chiscmmatch}) with
$a=-(H_I\sinh (\eta))^{-1}$. Choosing $N_*=H_I^{-1}$, we have
\beq
\varphi_*= \sinh (-\eta)\left(1-\epsilon -{1\over  \tanh(\eta)}
\right)e^{(1-\epsilon)\eta} \ .
\label{ucsmCDInf}
\eeq
In the early time limit ($\eta\to -\infty$),
$\varphi_*\simeq \frac{2-\epsilon}{2} e^{-\epsilon\eta}$.
Therefore, the two-point function is obtained
by replacing the factor $e^{-\epsilon\eta}$ in \eqref{phiphi2}
by $\frac{2}{2-\epsilon}\varphi_*$.

Near the end of inflation ($\eta\to -0$),
\eqref{ucsmCDInf} approaches a constant $\varphi_*\simeq 1$.
The two-point function in this limit (and in the
small $\epsilon$ limit) becomes\footnote{We have kept
$\epsilon$ in the exponent of the factor $(H_I/H_A)^{2\epsilon}$,
since we do not know the magnitude of $(H_I/H_A)$.
We have also kept the factor $(1+e^{2X_0})^{-2\epsilon}$
in (\ref{cstar}).}
\beq
 \langle \phi\phi\rangle^{\rm s.c.m.}
 =c_*\,  {H_{A}^4\over m_{A}^2}
 \left({H_I\over H_A}\right)^{2\epsilon} 
 \label{phiphi3}
\eeq
with an $X_0$-dependent constant
\beq
 c_*
= {1\over 2\pi^2}~
  {3\over 2-(2+\sech^2 X_0)\tanh X_0}~
  {1\over (1+e^{2X_0})^{2\epsilon}}
  \ .
\label{cstar}
\eeq
This has been obtained by using the expression (\ref{phiphi2})
for $\epsilon$ in the first-order perturbation theory.

In the RD era, $v_*$ is given by (\ref{chiscmRD}). Using
(\ref{aInfRDMD}) and (\ref{eta1alphaHI}), and setting $N_*=H_I^{-1}$ again,
$\varphi_*=v_*/a$ becomes
\beq
\varphi_* \simeq
\frac{\sinh\left((1-\epsilon)\eta\right)}{(1-\epsilon)\sinh(\eta)} \ ,
\label{ucsmRD}
\eeq
where the terms in
(\ref{chiscmRD}) of order ${\cal O}(\epsilon){\cal O}(\eta_2^3)$
are ignored. In the RD era, the conformal time
takes values between $\eta_2$ and $\eta_3$, and thus it is
tiny, $\eta \ll 1$ (See Section~\ref{subsec:scalefactor}).
Thus, the value of
$\varphi_* = 1-\frac{1}{6}\epsilon(2-\epsilon)\eta^2
+{\cal O}(\epsilon){\cal O}(\eta^4)$ does not change much,
and the two-point function (\ref{phiphi3}) receives only small corrections of
order ${\cal O}(\epsilon){\cal O}(\eta^2)$.

In the MD era, $\varphi_*=v_*/a$ can be found from (\ref{chiscmMD}),
(\ref{aInfRDMD}), (\ref{eta1alphaHI}) and (\ref{eta3alphabeta}),
\beqa
\varphi_* &\simeq&
\frac{3}{(8(1-\epsilon)^2-2)(1-\epsilon)^2}  \n
&& \frac{1}{\sinh^2(\eta/2)} \left(2(1-\epsilon)\cosh((1-\epsilon)\eta)
-\frac{\sinh((1-\epsilon)\eta)}{\tanh(\eta/2)}\right)  \ ,
\label{ucsmMD}
\eeqa
where again we neglect the terms  of order
${\cal O}(\epsilon) {\cal O}(\eta_3)$, ${\cal O}(\epsilon) {\cal O}(\eta_3^6)$
and ${\cal O}(\epsilon) {\cal O}(\eta_2^3)$
in  (\ref{chiscmMD}). As in the RD era, the conformal time stays
small in the MD era.
Thus, the value of the wave function $\varphi_*$
does not have significant change (see (\ref{phiMDexp})),
and the two-point function remains to be (\ref{phiphi3})
except for small corrections of order ${\cal O}(\epsilon){\cal O}(\eta^2)$.

Therefore, even though the non-zero $\epsilon$ introduces some
time dependence for the wave function, we can conclude that
this effect is not large for small $\epsilon$ and $\eta$.

\subsubsection{The $\epsilon=0$ approximation}
\label{subsec:epsilon0}

In the periods (ii) and (iii), the term $2\epsilon-\epsilon^2$
is the least relevant, and Eq.~(\ref{eqetascm}) is
approximated by neglecting  this term.
In terms of $\varphi_*=v_*/a$, this equation is rewritten as
\beq
\left[\partial_t^2+3H \partial_t+m_0^2 \right]\varphi_*(t)=0 \ ,
\label{eq_phi_t}
\eeq
where we have used the physical time $t$ instead of the conformal time $\eta$
(where $dt/d\eta=a$),
and $H=\partial_t a/a$ is the Hubble parameter.
This equation has the same form as the zero-modes
in the flat-space case.

Throughout the RD and MD eras (until today and perhaps much later),
the spatial curvature can be
neglected, as one can see from (\ref{NcInfRDMD}).
In the case of spatially flat universe,
the Hubble parameter behaves as $H \propto t^{-1}$.
For $H=p/t$ with $p$ being a constant parameter, the solutions (\ref{eq_phi_t})
are expressed in terms of the Bessel function as
\beq
\varphi_*=(m_0t)^{-\nu} \left( F~J_\nu(m_0t) +G~Y_\nu(m_0t) \right) \ ,
\label{phiBessel}
\eeq
where
\beq
\nu=\frac{3}{2} p-\frac{1}{2} \ ,
\eeq
and $F$ and $G$ are arbitrary constants.
In the MD period, $p=2/3$ and $\nu=1/2$. Then (\ref{phiBessel}) becomes
\beq
\varphi_*= (m_0t)^{-1} \left( F \sin (m_0t) +G \cos (m_0t) \right) \ .
\eeq
By smoothly connecting it to the frozen wave function
in the previous subsection, we find
\beq
\varphi_*\simeq
(m_0t)^{-1}  \sin (m_0t) \ .
\label{phiMDflat}
\eeq
In the period (ii),  $m_0 t\lesssim 1$ (or $m_0 H^{-1}\lesssim 1$),
and the wave function (\ref{phiMDflat}) is frozen.
In the period (iii), $m_0 t\gtrsim 1$ (or $m_0 H^{-1}\gtrsim 1$),
and then the wave function starts oscillating.
Note that at later times  $\eta \gtrsim 1$,
we have to take spatial curvature into account,
which will modify the wave function (\ref{phiBessel}).

\subsection{Time evolution of energy density and pressure}
\label{subsec:energydensity}

Let us now consider the expectation value of the energy-momentum
tensor. First note that the continuous modes are expected to give
negligible contributions. As mentioned in Section~\ref{sec-EMT},
the wave functions for the continuous modes decreases to
order $H_{I}$ by the end of the curvature domination, due
to the time dependence $e^{-\eta}$,
while the supercurvature mode decays only as $e^{-\epsilon\eta}$.
The continuous modes do not receive the enhancement factor
$\epsilon^{-1/2}$ either.
Thus, they are smaller by
a factor $\frac{m_A}{H_A}\left(\frac{H_{I}}{H_{A}}\right)^{1-\epsilon}$
relative to the supercurvature mode.

One may worry about the divergence in the coincident-point
limit. Such divergence appears in the continuous modes
where the renormalization/regularization is needed to introduce
the counter terms to cancel divergent pieces, resulting in the finite
terms in the energy-momentum tensors. Nevertheless, they can be
safely ignored  due to the fact that these terms
will be a combination of curvature tensors because the
renormalization is done in a local and covariant manner.
In the present universe, such  terms in the energy-momentum
tensor will be of order $H_{0}^{4}$, which is smaller than
the contribution from the supercurvature mode.

Thus, we concentrate on the contribution from the
supercurvature mode.
We will estimate
the energy density (\ref{rho1})
and pressure (\ref{p1}),
by comparing the magnitude of each term.
Note that the spatial-derivative term is
$\langle (\nabla\phi)^2 \rangle
= -\langle \phi \nabla^2 \phi \rangle
=(2\epsilon-\epsilon^2)\langle \phi  \phi \rangle
={\cal O}(\epsilon)$.

In the period (i),
the time-derivative term is
$\langle\varphi_*'^2\rangle={\cal O}(\epsilon^2)$,
since the wave function is almost frozen and
$\varphi_*'={\cal O}(\epsilon)$.
Hence, the spatial-derivative term is dominant in (\ref{rho1})
and (\ref{p1}), and then
\beqa
\rho &\simeq &
c_* \frac{H_A^4}{m_A^2}
\left(\frac{H_I}{H_A}\right)^{2\epsilon}
\frac{\epsilon}{a^2}
\sim H_A^2 \left(\frac{H_I}{H_A}\right)^{2\epsilon}
\frac{1}{a^2}  \ ,
\label{rho_spatialderiv}\\
p &\simeq & -\frac{1}{3}\rho \ ,
\label{p_spatialderiv}
\eeqa
where $c_*$ is given in (\ref{cstar}).
The equation of state is $w\simeq -1/3$.

In the period (ii), the wave function is almost frozen,
and the time-derivative term in (\ref{rho1})
and (\ref{p1}) is smaller than the mass term by
a factor of order $(m_0t)^2$ (or $(m_0/H)^2$).
From (\ref{phiMDflat}), we see
$\varphi_* \simeq 1-(m_0t)^2/6$ and
$\varphi'_*/a =\partial_t\varphi_*\simeq -(m_0t)/3 \cdot m_0\varphi_*$.
Then, the mass term is dominant in (\ref{rho1}) and (\ref{p1}),
and we find
\beqa
\rho &\simeq&  \frac{1}{2}c_* m_0^2
\frac{H_A^4}{m_A^2} \left(\frac{H_I}{H_A}\right)^{2\epsilon}
\label{rhoDElike} \ , \\
p &\simeq& -\rho \ .
\label{pDElike}
\eeqa
The equation of state is $w\simeq -1$, and it gives a dark energy
candidate.

The wave function (\ref{phiMDflat}) is valid over
the periods (ii) and (iii). With this wave function,
the energy density (\ref{rho1}) and pressure (\ref{p1})
are estimated as
\beqa
\rho &\simeq&
c_* \frac{H_A^4}{m_A^2} \left(\frac{H_I}{H_A}\right)^{2\epsilon}
\frac{m_0^2}{2(m_0t)^2} \n
&&\cdot\left[1-\frac{1}{m_0t} \sin (2m_0t)
+ \frac{1}{2(m_0t)^2}\left(1-\cos (2m_0t)\right) \right] \ ,
\label{rho_oscdec}
\eeqa
\beqa
p &\simeq&
c_* \frac{H_A^4}{m_A^2} \left(\frac{H_I}{H_A}\right)^{2\epsilon}
\frac{m_0^2}{2(m_0t)^2} \n
&&\cdot\left[\cos (2m_0t)-\frac{1}{m_0t} \sin (2m_0t)
+ \frac{1}{2(m_0t)^2}\left(1-\cos (2m_0t)\right) \right] \ .
\label{p_oscdec}
\eeqa
They can be rewritten as a function of $m_0/H$ instead of $m_0t$,
with the relation between time and the Hubble
parameter $t={2\over 3}H^{-1}$ in the MD era.
In the period (ii),
 $m_0 t\lesssim 1$ (or $m_0/H\lesssim 1$),
and (\ref{rho_oscdec}) and (\ref{p_oscdec}) reduce to
(\ref{rhoDElike}) and (\ref{pDElike}) respectively,
as we can see by considering the leading terms in
the expansion in $m_0t$.
In the period (iii),
$m_0 t \gtrsim1$ ($m_0/H\gtrsim 1$),
and the energy density and pressure show the oscillating and decreasing behavior.

We now summarize the time evolution of the energy density and pressure.
\begin{enumerate}
\item[(i)] When $m_{0}^2a^2<2\epsilon-\epsilon^2$ is satisfied,
the energy density is given by (\ref{rho_spatialderiv}),
and the equation of state is $w=-1/3$.
\item[(ii)] When $2\epsilon-\epsilon^2<m_{0}^2a^2<a''/a$, assumed to be satisfied in the present universe,
the energy density is given by (\ref{rhoDElike}), leading to $w=-1$.
 \item[(iii)] When $a''/a<m_{0}^2a^2$  occurs
as $m_0/H\gtrsim 1$, the energy density and pressure oscillate
as in (\ref{rho_oscdec}) and (\ref{p_oscdec}).
If we take an average over the period of oscillation (using the Virial theorem),
we obtain $w=0$. Thus, the energy density will decay as fast
as the energy density of matter\footnote{If dark energy eventually decays as
assumed in this paper, the universe
will be curvature dominated again at very late times,
since the energy density of curvature decays more slowly than the one for matter.
To study that regime, we will have to use the wave function of the scalar field on
 the curvature-dominated background, and also may have to consider the back
reaction from the scalar field to the geometry.}.
\end{enumerate}

We finally give some comments about the continuous modes.
The wave functions with $k\gtrsim1$ (or $k_{\rm phys}=k/a \gtrsim R_c^{-1}$)
do not have much change from those in the flat-space geometry,
studied in \cite{AIS,AI}, where
modes with $k \sim 1/\eta$ (or $k_{\rm phys} \sim H$)
mainly contribute to the derivative terms of the
energy-momentum tensor. Then,
as long as $\eta \lesssim 1$ (or $H^{-1} \lesssim R_c$)
(i.e., throughout the RD and MD eras
until today and later times),
the results for the flat-space geometry obtained in \cite{AIS,AI}
can be applied:
\beq
\rho \sim H_I^2 H^2 
\label{rho_conmod}
\eeq
with the equation of state $w=1/3$ in the RD era
and $w=0$ in the MD era.
At earlier times, (\ref{rho_conmod}) is dominant over the
supercurvature-mode contributions.
The transition from this epoch to the period (i) occurs
when (\ref{rho_spatialderiv}) dominates  over (\ref{rho_conmod}).
Neglecting the numerical factors, and solving
$H_A^2a^{-2}=H_I^2H^2$ with the use of $H^2=\alpha^2 a^{-4}$ in the RD era
and $H^2=\beta a^{-3}$ in the MD era, we find the transition occurs
when
\beq
H\sim\frac{1}{\alpha H_0}\left(\frac{H_A}{H_I}\right)^2H_0
\eeq
if it happens in the RD era, and
\beq
H\sim\frac{1}{\beta H_0}\left(\frac{H_A}{H_I}\right)^3H_0
\eeq
if it happens in the MD era.
The prefactors are constrained by
$1/(\alpha H_0)\lesssim 1/2$ and $1/(\beta H_0)\lesssim 10^{-3}$
using (\ref{albeest}) and (\ref{betaest}).

\subsection{Vacuum energy as dark energy}
\label{subsec:vaceneDE}

To interpret the vacuum energy in the period (ii) as the present dark energy,
the following three conditions are necessary.
One is that the vacuum energy (\ref{rhoDElike}) has the same
order of magnitude
as dark energy, written explicitly as
\beqa
&&\frac{1}{2}c_*m_0^2\frac{H_A^4}{m_A^2}
\left(\frac{H_I}{H_A}\right)^{2\epsilon}
\simeq 3\Omega_\Lambda H_0^2 M_p^2  \ , \\
&\Leftrightarrow &
\frac{m_0}{H_0}~\frac{H_A}{m_A}~\frac{H_A}{M_p} \simeq
\left(\frac{6\Omega_\Lambda}{c_*}\right)^{1/2}
\left(\frac{H_A}{H_I}\right)^{\epsilon} \ ,
\label{cond1}
\eeqa
with $\Omega_\Lambda\sim 0.7$.
The second and third conditions are that the present moment is in
the period (ii):
\beq
2\epsilon -\epsilon^2 < m_0^2a_0^2 <  \frac{a''_0}{a_0}  \ .
\eeq
With the scale factor (\ref{aInfRDMD}) in the MD era, they become
\beq
2 c_\epsilon \left(\frac{m_A}{H_A}\right)^2
< \left(\frac{m_0}{H_0}\right)^2 \coth^2(\eta_0/2)
< \frac{1}{2\sinh^2(\eta_0/2)}+1 \ ,
\label{inequalities}
\eeq
where we assume $\epsilon\ll 1$, and
$c_\epsilon$ is a constant of order unity,
defined by $\epsilon=c_\epsilon m_A^{2} H_A^{-2}$.
From (\ref{epsilonpert}),
\beq
c_\epsilon = {2-(2+\sech^2 X_0)\tanh X_0\over 3\, (2+\sech^2 X_0-2\tanh X_0)} \ .
\eeq
To obtain \eqref{inequalities},
(\ref{NcInfRDMD}) and (\ref{PotentialInfRDMD}) have been applied.
With (\ref{eta0Nc0}),
 \eqref{inequalities} can be
further rewritten as
\beq
2 c_\epsilon \left(\frac{m_A}{H_A}\right)^2
< \left(\frac{m_0}{H_0}\right)^2 r_{c0}^2
< \frac{1}{2} r_{c0}^2 \ .
\label{cond2}
\eeq

Let us heuristically explain the conditions
in \eqref{cond2}.  The first inequality requires that
the eigenvalue of Laplacian is smaller than the mass.
Laplacian is associated with the inverse
scale factor squared, and
$\epsilon a_0^{-2}< m_0^2$. Recalling the fact that the
present curvature radius is equal to $a_0$, we can rewrite the
first inequality using $a_{0}=r_{c0}H_{0}^{-1}$.
The second inequality is the condition for
non-oscillation of the wave function, and gives
$m_{0} < H_{0}$.

The conditions (\ref{cond1}) and (\ref{cond2}) can be
satisfied by physically acceptable values of the
parameters\footnote{Here, the right-hand
side of (\ref{cond1}) and $c_\epsilon$ in (\ref{cond2})
are considered to be of order unity.}, for example,
\beq
H_A \sim M_p \ , \ \
\frac{m_A}{H_A} \sim \frac{m_0}{H_0} < 1 \ . \ \
r_{c0} > 1 \,
\eeq
One could also consider the case that the second
inequality of \eqref{cond2} is barely satisfied,
$m_{0}\sim H_{0}$, leading to
\begin{equation}
{H_A\over M_P}\sim {m_A\over H_A} < r_{c0} \ ,
\end{equation}
which means $H_A$ is the geometric mean of
$M_P$ and $m_A$.

The transition from the period (i) with $w=-1/3$
to the period (ii) with  $w=-1$ is an interesting
signature of our mechanism for the realization of dark energy.
The transition occurs at the time
when $2\epsilon-\epsilon^2 = m_0^2a^2$,
i.e., at
\beq
1+z=\frac{a_0}{a}
\simeq \frac{1}{\sqrt{2\epsilon}} \ \frac{m_0}{H_0} \ r_{c0} \ .
\eeq

Let us compute $w(z)$ as a function of the redshift $z$,
assuming that the time derivative terms
in $\rho$ and $p$ can be ignored. This should be a good
approximation in the period (i) and (ii), though the
precision of the approximation will depend on the
choice of the parameters.
In this case, we get a very simple
expression,
\begin{equation}
 w(z)= - {{2\over 3}\epsilon+m_0^2 a^2\over
  2\epsilon+m_0^2 a^2}
  =- {1+{2\over 3}\tilde{\epsilon}(1+z)^2
  \over 1+2\tilde{\epsilon}(1+z)^2},
\label{wz}
\end{equation}
where the above $\tilde{\epsilon}$ is defined as
\begin{equation}
 \tilde{\epsilon}={\epsilon\over (m_0/H_0)^2 r_{c0}^2}.
\end{equation}
Here we have used
  $a=r_{c0}H_0^{-1}(1+z)^{-1}$ to obtain the right hand side
  of \eqref{wz}.
  When this approximation is valid, $w(z)$ depends only
 on $\tilde\epsilon$, as it can be seen in \eqref{wz}.
 The deviation of the present equation of state $w_0=w(0)$
 from $-1$ is
 \begin{equation}
  w_0+1={{4\over 3}\tilde{\epsilon}\over
   1+2\tilde{\epsilon}}\approx {4\over 3}\tilde{\epsilon}.
 \end{equation}
 The last approximation is for $\tilde{\epsilon}\ll 1$
 to be satisfied in most cases of our study here.
 The derivative of $w$ with respect to the scale factor
 (evaluated at present), which is sometimes called $w_1$,
 (see e.g., \cite{PlanckCosm}), is
  \begin{equation}
   w_1=-a{dw\over da}\Big|_{a=a_0}={8\over 3}
	{\tilde{\epsilon}\over (1+2\tilde{\epsilon})^2}
	\approx {8\over 3}\tilde{\epsilon}.
  \end{equation}
  Thus, as long as our approximation of 
  neglecting the time derivative and taking 
  $\tilde{\epsilon}\ll 1$ is valid,
  there is a simple relation
  between $w_0$ and $w_1$, namely, $w_1=2(w_0+1)$.

\begin{figure}[!htb]
\center
\includegraphics [scale=1.1]
 {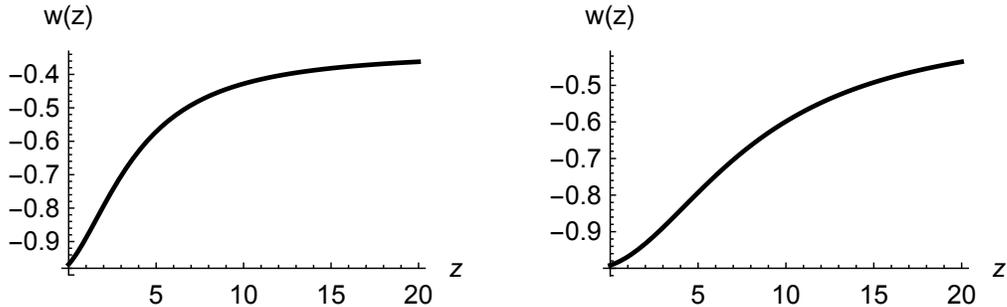}
 \caption{The equation of state $w(z)$ as a function of the
 redshift $z$. The parameters for the left panel are
 $\epsilon=0.1$, $m_0/H_0=0.1$, $r_{c0}=20$, for which
the present equation of state is $w_0=-0.968$. Those for the right panel are
 $\epsilon=0.1$, $m_0/H_0=0.1$, $r_{c0}=40$, for which
 $w_0=-0.992$. Note that these are preliminary results,
 obtained by ignoring
 the time derivative terms in $\rho$ and $p$ (though this
 approximation can be justified for these choices of parameters).}
 \label{Fig-wz}
\end{figure}

In Figure~\ref{Fig-wz}, we show the plot of $w(z)$ for two
choices of the set of
parameters $\epsilon$, $m_0/H_0$, $r_{c0}$.
These parameters have been chosen so that the time derivative
terms in $\rho$ and $p$ can be safely ignored. The validity of
 this approximation was confirmed by 
studying the time dependence of the solutions
obtained in Sections 5.2.2 and 5.2.3, where the dominant term
of $2\epsilon$ and $m_0^2 a^2$ have been kept to
solve the equation of motion. It is an important subject
for future investigations to obtain $w(z)$ for more
general choices of parameters without introducing an approximation.

\section{Conclusions}

Let us summarize our results.
We have calculated the vacuum expectation value of the
energy-momentum tensor for a scalar field in an open
universe created by bubble nucleation. We pay
particular attention to the contribution from the
supercurvature mode, a non-normalizable mode on $H^3$,
which appears when the mass $m_A$ in the ancestor vacuum
is small enough. The vacuum expectation value of
the energy-momentum tensor in the early-time limit is
obtained by the Euclidean prescription. Then,
its time evolution is studied using the equation
of motion for the scalar field.
The supercurvature mode decays more slowly
than the continuous modes, thus it gives the most
important contribution at late times. We have shown
that the vacuum energy for a minimally coupled
scalar field can be
regarded as dark energy. For this interpretation, it is needed
that there is a field with
mass (in the true vacuum) $m_0$ of order the present
Hubble parameter $H_0$ or smaller, and the ratios of
$m_0$, $m_A$ and the Hubble parameter in the ancestor vacuum
$H_A$ satisfy certain inequalities. The latter condition
does not seem difficult to satisfy.
As long as $m_0\lesssim H$,
the field value is essentially frozen (though there
is weak time dependence due to non-zero $m_A$).
In the future, when the Hubble parameter decreases so that
$m_0 \gtrsim H$, the energy density decays.

A nice point about our analysis is that the main result is free of
theoretical uncertainties in the following senses.
First, our result
does not depend on the renormalization prescription
for the UV divergence, since the supercurvature mode is
non-singular in the coincident-point limit. Vacuum
energy is often considered to be ambiguous due to the
UV divergence, but we believe our finite result has the
intrinsic physical meaning.
Second, the change of mass from $m_A$ to $m_0$, which
is assumed to occur during tunneling, does not
give rise to complicated non-equilibrium processes.
This change occurs in the spatial direction in Region~III
in Fig.~\ref{Fig-Penrose}, with its effect essentially
encoded in the initial condition in the FLRW universe,
such as the position of
the pole for the bound states in the Euclidean problem.
Third, we treat the field $\phi$ fully quantum
mechanically, and do not have to assign a
random classical value for the field $\phi$.
In the study of axions, one sometimes
has to set the misalignment angle by hand, but we
do not have that kind of ambiguity.

We do not expect the fields other than scalars to give
contributions of the type studied in this paper.
Massless vectors (spin 1) are Weyl invariant in
3+1 dimensions, and so are massless spinors (spin 1/2)
in any dimensions. A supercurvature mode appears when
there is a bound state in
the complete set in the $X$ direction
(see Fig.~\ref{Fig-Penrose}).  For the scalar case,
the presence of the bound state is due to the non-trivial
potential $a''/a$ in the equation of motion arising from the
coupling with the curved background. Weyl-invariant fields
have the same equation of motion as in the flat spacetime, and
there is no supercurvature
mode\footnote{See \cite{YamauchiMukohyama} for recent
work on the absence of supercurvature modes for
vector fields.}.
Gravitons (spin 2) behave
similarly to the massless scalar, but the counterpart of
 the $k=i$ supercurvature mode
in the scalar case
is known to be pure gauge
in the spin 2 case~\cite{SasakiTanakaGraviton}.
Furthermore,
we cannot give mass to gravitons, so there is
no mechanism for generating the vacuum
energy from the mass term.
The remaining
field is gravitino (spin 3/2).
This field has not been studied in the context of
bubble nucleation, and it is an interesting question
on how the correlation functions and the
vacuum energy of gravitino  behave. However, it is not likely that
the mass of gravitino is of order $H_0$,
since its mass is related to the
scale of supersymmetry breaking, and as we know,
there is no supersymmetry at such a low energy scale.

In this paper, we have not discussed the origin of
the scalar field $\phi$ with mass of order
$m_{0}\sim H_0\sim 10^{-33}{\rm eV}$.
We expect it to be one of the many axion-like
particles that have been proposed to
exist in superstring theory according to the idea of
``string axiverse''~\cite{Axiverse}.
String axiverse have been studied in
the framework of
type IIB string theory~\cite{Cicoli1, Cicoli2}
and M-theory~\cite{Acharya}. These studies will serve
as a starting point for an explicit construction of
the field $\phi$ considered in this paper.
One interesting possibility is that the field $\phi$
simultaneously gives dark energy and dark matter.
To serve as dark matter, axions should have
mass $m_{\rm DM}\gtrsim 10^{-22}{\rm eV}$
(See e.g., \cite{Marsh, Witten, MarshReview})\footnote{%
Dark matter with an extremely low mass
$m_{\rm DM}\sim 10^{-22}{\rm eV}$, sometimes called
``fuzzy'' dark matter having de Broglie
wavelength $\lambda\sim 1 {\rm kpc}$, has attracted
attention recently. It is considered as a possible resolution
of the apparent inconsistencies of
the cold dark matter (CDM) model with
the observations of galaxies at length scales below
$10 {\rm kpc}$ (overabundance of the structure on small scales
in CDM)~\cite{Marsh, Witten, MarshReview}.}.
If the mechanism for generating the mass $m_{\rm DM}$ is
given by the local dynamics inside our universe, it could be that
while the continuous modes get mass $m_{\rm DM}$,
the supercurvature mode is unaffected
to still have mass $m_{0}$, because the latter mode is essentially
determined in the ancestor vacuum. This point needs
further study.

It is highly important to test observationally
whether dark energy has been produced by our mechanism
or not. According to our proposal, the equation of state
of dark energy will deviate from $w=-1$ as we go back
in the past, and will approach $w=-1/3$. The transition
occurs when the spatial derivative term (of order $\epsilon$)
becomes dominant over the mass term $m_0^2a^2$ in the energy
density \eqref{rho1} and pressure \eqref{p1}.
The time of this transition depends on the parameters
$r_{c0}$ (the ratio of the curvature radius to $H_0^{-1}$),
$m_0/H_0$ and $\epsilon$, but there is a characteristic
behavior of $w(z)$, which seems to be rather general
as mentioned at the end of Section~5.
The observational
determination of the dark energy equation of state $w(z)$
as a function of $z$ would be a great
 challenge\footnote{See \cite{Riess} for a review
 on the observational probes of cosmic acceleration.},
and it would be especially difficult to obtain
 $w(z)$ for high redshift, since dark energy will be
 less and less important than the energy density of matter at early times. Nevertheless,
confirmation (or rejection) of the pattern
like the one shown in Figure~\ref{Fig-wz}
might be within reach of the observations in the
near future.
One of such observational projects
would be the multi radio telescope,
  Square Kilometre Array
  (SKA)~\cite{SKA},
  which is expected to
 deliver precise cosmological measurements
through the survey of a large number of distant
 galaxies using the 21 cm hydrogen
 line~\cite{SKAJapan, Kohri}.
 Giving detailed theoretical
 predictions for $w(z)$
 for comparison with observations is an important
subject for future studies.

\section*{Acknowledgments}
We thank Daisuke Yamauchi for helpful comments on the first version.
This work is supported in part by Grants-in-Aid for Scientific
Research (Nos.\ 16K05329, 24540279) from
the Japan Society for the Promotion of Science.
DSL and CPY are supported in part by the Ministry of
Science and Technology, Taiwan.

\section*{Appendix}
\appendix
\section{Harmonics on $H^3$}
\label{sec-modes}

The eigenvalue equation in (\ref{LapfkR}) can be written down
explicitly as
\begin{equation}
 \left[{1\over \sinh^2 R}\partial_R \left(\sinh^2 R\partial_R\right)
+{\nabla^2_{S^2} \over \sinh^2R} \right]f^{(lm)}_k(R, \Omega_2)=
-(k^2+1)f^{(lm)}_k(R, \Omega_2) \ ,
\end{equation}
where $\nabla^2_{S^2}$ is the Laplacian on $S^2$.
Solutions are given as
 \begin{equation}
 \label{mode1}
 f^{(lm)}_k(R,\Omega_2)=\Pi^{(kl)}(R)Y^{(lm)}(\Omega_2) \ ,
 \end{equation}
where $Y^{(lm)}(\Omega_2)$ is the spherical harmonics on $S^2$,
which satisfy $\nabla^2_{S^2} Y^{(lm)} = -l(l+1) Y^{(lm)}$.
The function $\Pi^{(kl)}(R)$ with the normalization condition
(\ref{norm}) can be written as (for $0<k<\infty$),
  \be
\Pi^{(kl)}(R)=\sqrt{\frac2{\pi}}\left(\prod_{n=0}^{l}(k^2+n^2)\right)^{-1/2}\sinh^l
 R \left(\frac{-1}{\sinh R}\frac{d}{dR}\right)^{l+1}\cos(k R)
 \ee

\section{Calculation of $\langle \phi^2\rangle$ in pure de Sitter space}
\label{sec-puredS}

 In this Appendix, we will calculate
 $\langle \phi^2\rangle$ in pure de Sitter space
 in the massless limit.
 The derivation is somewhat simpler than the one presented
 in Section~4, where the result was obtained by
 taking the limit of small bubble.
The calculation will be performed in the Euclidean space.
 We will see that the contribution from the supercurvature mode
 (i.e., the
 bound state in the Euclidean problem) reproduces
 the result obtained by the standard techniques.

 We start from the expression \eqref{Euclideancorrelator0}
 for the Euclidean two-point function
 $\langle \chi(X, \theta)\chi (X', 0)\rangle$. We take
 the bound state contribution, divide it by the scale factors
 to obtain
 $\langle \phi(X, \theta)\phi (X', 0)\rangle$, take the
 massless
 ($\epsilon\to 0$)
 and coincident-point
 ($X\to X'$, $\theta\to 0$)
 limit, and obtain
 \begin{equation}
  \langle \phi^2\rangle = \lim_{\epsilon\to 0}\lim_{X'\to X}
  \lim_{\theta\to 0}
  {u_{k_B}^{B}(X)
  u_{k_B}^{B*}(X')\over a(X)a(X')}G_{k_B}(\theta),
\label{phiphiapp}
 \end{equation}
where $k_B=(1-\epsilon)i$ denotes the position of the bound state
 pole. %

 The function $G_k(\theta)$ is the Green's function
  on $S^3$ with the effective mass $k^2+1$. It is the
  solution of \eqref{Gkthetaeq}, given explicitly
  by \eqref{Gktheta}.
 It is singular in the $\epsilon\to 0$ ($k\to i$) limit.
 We will be interested in the leading singularity,
 proportional to $1/\epsilon$. Let us rewrite \eqref{Gktheta}
as
  \begin{equation}
   G_k(\theta)={1\over 4\pi\sinh k\pi \sin\theta}
	\left( \sinh (k\pi)\cosh(k\theta)
	 -\cosh(k\pi)\sinh(k\theta)\right).
  \end{equation}
  The first term is singular in the $\theta\to 0$ limit, but
  the coefficient of this singularity is independent of $\epsilon$.
  This term represents
  the UV divergence which exists generally in quantum field theory,
  thus can be
  discarded\footnote{If we analytically continue to the open
  FLRW universe following the prescription described in
  Section~3.2, this UV divergence dissapears from the
  supercurvature mode. See \eqref{chichiNNM}.}.
  Renormalization of this
  term will give rise to finite terms expressed
  in terms of local curvature tensors, but we will
  not consider such terms here. The second term
  is regular in the $\theta\to 0$ limit, but diverges as
  $1/\epsilon$ in the $\epsilon \to 0$ limit. This term
  gives the dominant contribution that we are interested
  in,
 \begin{equation}
  \lim_{\epsilon\to 0} \lim_{\theta\to 0}
   G_{k_B}(\theta)= {1\over 4\pi^2\epsilon}.
\label{leading}
 \end{equation}

 To compute the factors multiplying \eqref{leading}
 to obtain \eqref{phiphiapp},
 we can substitute $k=i$ for the bound
 state wave function $u_{k=i}(X)$, since
 this function is regular in the $\epsilon\to 0$ limit.
 The wave function with $k=i$ is proportional to the
 scale factor $a(X)$, as explained in \eqref{uB}.
 For pure de Sitter space, we have $a(X)=H_{A}^{-1}/\cosh X$,
 and
 \begin{equation}
  u_{k=i}^{B}(X)={\cal N}{1\over \cosh X},
 \end{equation}
 where
 \begin{equation}
  {\cal N}=\left(\int_{-\infty}^{\infty}dX {1\over \cosh^2 X}
			\right)^{-1/2}={1\over \sqrt{2}}.
 \end{equation}

 The parameter $\epsilon$ can be obtained by the first-order
 perturbation ($2\epsilon$ being the eigen-energy)
 as explained in Section 3,
 \begin{equation}
  \epsilon={1\over 2}\int_{-\infty}^{\infty}dX m_A^2 a^2(X)
   \left(u_{k=i}^{B}(X)\right)^2=
 {1\over 2}  m_A^2H_A^{-2}{\cal N}^2\int_{-\infty}^{\infty}dX
   {1\over \cosh^4 X}={1\over 3} m_A^2H_A^{-2}.
 \end{equation}
 Putting these factors together, the expectation value
 \eqref{phiphiapp} becomes
 \begin{equation}
\langle \phi^2\rangle ={3\over 8\pi^2}{H_A^{4}\over m_A^2},
 \end{equation}
 which agrees with the known value
 (apart from the finite contribution due to
renormalization) in pure de Sitter space
  (see e.g., \cite{BirrellDavies, Linde}).

\section{Scale factor for open universe with both matter and radiation}
\label{sec-radmatt}

We consider the cases where the energy density contains
both radiation and matter component,
\begin{equation}
\rho=3M_p^2\left({\alpha^2\over a^4}+{\beta\over a^3}\right) \ .
\label{rhoMDRD}
\end{equation}
Solutions of the Friedmann equation \eqref{Friedmann} with
this energy density are
\begin{equation}
 a(\eta)= -\frac{\beta}{2}
\pm \left(\alpha^2-\frac{\beta^2}{4}\right)^{1/2} \sinh (\eta)
\end{equation}
for $\alpha^2-\frac{\beta^2}{4}>0$, and
\begin{equation}
 a(\eta)= -\frac{\beta}{2}
\pm \left(\frac{\beta^2}{4}-\alpha^2\right)^{1/2} \cosh (\eta)
\end{equation}
for $\frac{\beta^2}{4}-\alpha^2>0$,
up to a shift in $\eta$.
The scale factors in the RD and the MD eras,
given in (\ref{aInfRDMD}),
are reproduced from the above solutions
by setting $\beta=0$ (RD) and $\alpha=0$ (MD),
respectively.

\section{Matching of the wave functions}
\label{sec-matwf}

Requiring the continuity conditions of the wave functions (\ref{chiscmmatch})
across the eras,
the coefficients $A$, $B$, $C$ and $D$ are determined as
\beq
\begin{pmatrix} A \cr B \end{pmatrix}
=\frac{N_*}{2(1-\epsilon) s_2^2}
\begin{pmatrix} \left[1+2(1-\epsilon)s_2c_2+2(1-\epsilon)^2s_2^2\right]
e^{-2(1-\epsilon)\eta_2} \cr
-1 \end{pmatrix}  \ ,
 \label{AB}
 \eeq
 with $s_2=\sinh(\eta_2)$, $c_2=\cosh(\eta_2)$, and
 \beqa
\begin{pmatrix} C \cr D \end{pmatrix}
&=&
\frac{1}{8(1-\epsilon)^2-2} \cdot \frac{1}{2(1-\epsilon) s_3^2}
\label{ABCD} \\
&&
\hspace*{-1.2cm}
\cdot
\begin{pmatrix}
[1+4(1-\epsilon)s_3c_3+8(1-\epsilon)^2s_3^2]e_3^{-1}
& e_3^{-3} \cr
-e_3^3
&-[1-4(1-\epsilon)s_3c_3+8(1-\epsilon)^2s_3^2] e_3
\end{pmatrix}
 \begin{pmatrix} A \cr B \end{pmatrix} \ ,
\nonumber
\eeqa
with $s_3=\sinh(\eta_3)$, $c_3=\cosh(\eta_3)$,
$e_3=e^{(1-\epsilon)\eta_3}$.

We are mostly interested in the case of small $\epsilon$
(i.e., $m_A\ll H_A$). Also note that
$\eta_2$ and $\eta_3$, which were determined in
Section~\ref{subsec:scalefactor}, are tiny numbers.
Expanding the upper element of (\ref{AB})
with respect to $\epsilon$ and $\eta_2$,
we obtain
\beq
A=\frac{N_*}{2(1-\epsilon) s_2^2}
\left[1+\frac{4}{3}(2\epsilon-3\epsilon^2+\epsilon^3)\eta_2^3
+{\cal O}(\epsilon){\cal O}(\eta_2^4) \right] \ .
\eeq
Then, by expanding the matrix elements of (\ref{ABCD})
with respect to $\epsilon$ and $\eta_3$, we find
\beqa
C+D&=&\frac{1}{8(1-\epsilon)^2-2}
\cdot\frac{N_*}{4(1-\epsilon)^2 s_2^2 s_3^2} \\
&&\cdot\left[\frac{16}{45}\epsilon(1-\epsilon)^2
(6-19\epsilon+16\epsilon^2-4\epsilon^3)\eta_3^6
+{\cal O}(\epsilon){\cal O}(\eta_3^8)
+{\cal O}(\epsilon){\cal O}(\eta_2^3)\right] \ , \nonumber
\eeqa
\beqa
C-D&=&\frac{1}{8(1-\epsilon)^2-2}
\cdot\frac{N_*}{4(1-\epsilon)^2 s_2^2 s_3^2} \label{CminD}\\
&&\hspace*{-1.2cm}\cdot\left[12\sinh(\eta_3)
+\epsilon\left(-12\eta_3+\frac{14}{3}\eta_3^3+{\cal O}(\eta_3^5)\right)
+{\cal O}(\epsilon^2){\cal O}(\eta_3^3)
+{\cal O}(\epsilon){\cal O}(\eta_2^3)\right] \ . \nonumber
\eeqa

Substituting these coefficients back into (\ref{chiscmmatch}),
we obtain the wave function
\beq
v_*=\frac{N_*}{2(1-\epsilon) s_2^2}
\left[2\sinh\left((1-\epsilon)\eta\right)
+{\cal O}(\epsilon){\cal O}(\eta_2^3)\right]
\label{chiscmRD}
\eeq
in the RD era, and
\beqa
v_*&=&\frac{1}{8(1-\epsilon)^2-2}
\cdot\frac{N_*}{4(1-\epsilon)^2 s_2^2 s_3^2}  \n
&&\hspace*{-1.2cm}\cdot\Biggl[\Bigl(12 s_3
+{\cal O}(\epsilon){\cal O}(\eta_3)+{\cal O}(\epsilon){\cal O}(\eta_2^3)\Bigr)
\left(2(1-\epsilon)\cosh((1-\epsilon)\eta)
-\frac{\sinh((1-\epsilon)\eta)}{\tanh(\eta/2)}\right) \n
&&\hspace*{-1.2cm}+{\cal O}(\epsilon){\cal O}(\eta_3^6)
+{\cal O}(\epsilon){\cal O}(\eta_2^3)\Biggr] \label{chiscmMD}
\eeqa
in the MD era.

We finally add a comment about the MD-era wave function $\varphi_*=v_*/a$,
given in (\ref{ucsmMD}).
Expanding (\ref{ucsmMD}) in terms of $\epsilon$ and $\eta$, we obtain
\beq
\varphi_*=\frac{1}{1-\epsilon}\left[1-\frac{1}{10}\epsilon(2-\epsilon)\eta^2
+{\cal O}(\epsilon){\cal O}(\eta^4)\right] \ .
\eeq
If we estimate the coefficient $C-D$ more precisely, taking into account
the term of order $\epsilon$ in (\ref{CminD}) as well,
we find
\beqa
\varphi_*&=&\left(1+\frac{5}{9}\epsilon\eta_3^2
+{\cal O}(\epsilon){\cal O}(\eta_3^4)+{\cal O}(\epsilon^2){\cal O}(\eta_3^2)
+{\cal O}(\epsilon){\cal O}(\eta_2^3)\right) \n
&&\cdot\left[1-\frac{1}{10}\epsilon(2-\epsilon)\eta^2
+{\cal O}(\epsilon){\cal O}(\eta^4)\right] \ .
\label{phiMDexp}
\eeqa


\end{document}